\RequirePackage{fix-cm}
\documentclass[smallextended]{svjour3}       
\smartqed  
\usepackage{graphicx}
\usepackage{amssymb}
\usepackage{amsmath}
\usepackage{amsfonts}
\usepackage{bm}
\usepackage{color}

\newcommand\etc{etc.}

\newcommand\eg{e.g.}
\newcommand\ie{i.e.}

\begin{document}

\title{Adaptive gradient-augmented level set method with multiresolution error estimation\thanks{This
        work was supported by the CRM--ISM Fellowship, NSERC Discovery and Discovery Accelerator Programs.}
}

\titlerunning{Adaptive gradient-augmented level set method}        

\author{Dmitry~Kolomenskiy \and Jean-Christophe~Nave \and Kai~Schneider}


\institute{D. Kolomenskiy \at
              CRM and The Department of Mathematics and Statistics, McGill University \\
              805 Sherbrooke W. Street \\
              Montreal, QC, Canada \quad H3A 0B9 \\
              Tel.: +1 514 398-3853\\
              \email{dkolom@gmail.com}           
           \and
           J.-C. Nave \at
              The Department of Mathematics and Statistics, McGill University \\
              805 Sherbrooke W. Street \\
              Montreal, QC, Canada \quad H3A 0B9
           \and
           K. Schneider \at
              M2P2--CNRS, Aix-Marseille Universit\'e \\
              39, rue Fr\'ed\'eric Joliot-Curie \\
              13453 Marseille Cedex 13, France
}

\date{Received: date / Accepted: date}

\maketitle

\begin{abstract}
A space-time adaptive scheme is presented for solving advection equations in two space dimensions.
The gradient-augmented level set method using a semi-Lagrangian formulation with backward time integration is coupled
with a point value multiresolution analysis using Hermite interpolation.
Thus locally refined dyadic spatial grids are introduced which are efficiently implemented with dynamic quadtree data structures.
For adaptive time integration, an embedded Runge--Kutta method is employed.
The precision of the new fully adaptive method is analysed and speed up of CPU time and memory compression
with respect to the uniform grid discretization are reported.
\end{abstract}

\section{Introduction}

In some advection dominated problems, the solution develops small-scale features but remains smooth during time evolution.
These problems can be solved efficiently using numerical methods based on high-order interpolation on fixed Eulerian grids \cite{Shu_Osher_1988,Liu_etal_1994}.
If the small-scale features are localized in some part of the computational domain,
its non-uniform partition with grid points clustered at the same part of the domain allows reducing the cost of computation without loosing accuracy.
However, if the location of these features changes in time,
the efficiency of the numerical method can be significantly improved by adapting the partition dynamically to the solution (see, \eg, \cite{Domingues_etal_2008} and references therein).

When applied to pure advection problems, Eulerian schemes require some stabilization which introduces numerical diffusion and thus pollutes the solution. Another drawback are small time steps imposed by the stability limit of explicitly discretized Eulerian schemes. 
Semi-Lagrangian schemes combine advantages of Eulerian schemes, such as connectivity of the grid, 
with those of Lagrangian schemes, especially that they have less demanding restrictions on the time step.
A review on semi-Lagrangian schemes introduced in the context of numerical weather prediction can be found, e.g., in \cite{Staniforth_Cote_1991}.
These schemes have also been used in the context of plasma physics for solving the Vlasov equation, e.g., \cite{Sonnendrucker_1999}.

In this paper, we present an adaptive method for the two-dimensional advection equation based on a semi-Lagrangian approach.
Advection problems are encountered for example in moving fronts for a given velocity field, or in transport of passive scalars modeling pollution or mixing in chemical engineering \cite{Ottino_1989}.
It can also be viewed as a simple model that partly describes other,
more complex problems, such as advection-reaction-diffusion, fluid flow, elasticity, \etc~
Therefore the proposed numerical method may be relevant to those problems as well.

We present a generalization of the gradient-augmented level set method \cite{Seibold_etal_2012,Nave_etal_2010,Chidyagwal_etal_2012} to adaptive discretization in space and in time.
There exists a large variety of approaches to introduce adaptivity
which differ in many aspects such as mesh topology, refinement criteria and data structure management.
A complete review is beyond the scope of the paper and
we mention exemplarily just a few.
Optimal grid adaptation based on a posteriori error estimators have been proposed in the context of finite element methods and are reviewed in \cite{Becker_Rannacher_2001,Verfuerth_2013}.
Cartesian grid methods have a long history, see, e.g., \cite{Aftosmis_1997},
and adaptive mesh refinement techniques \cite{Berger_Colella_1989,Berger_Oliger_1984,Martin_Colella_2000,Berger_LeVeque_1998,Ziegler_etal_2011} which are mostly based on Cartesian meshes are now well established to perform efficient simulations of engineering and science problems governed by conservation laws on massively parallel computers.
For a review on block-structured adaptive mesh refinement including implementation and application aspects we refer, e.g., to \cite{Deiterding_2011}.
An overview on different adaptive mesh refinement schemes including also multiresolution techniques can be found in the proceedings volume \cite{Plewa_etal_2005}.


In our approach, the refinement criterion uses an error estimate obtained from multiresolution analysis.
Such multiresolution based methods were first developed for conservation laws by Harten \cite{Harten_1995}.
Nowadays multiresolution techniques are known to yield an appropriate framework to construct fully adaptive schemes for hyperbolic conservation laws.
Extensions and further developments of Harten's original approach can be found, e.g., in \cite{Kaibara_Gomes_2001,Chiavassa_Donat_2001,Roussel_Schneider_2010}.
Recently fully adaptive multiresolution simulations of two-dimensional incompressible viscous flows have been proposed even on multicore architectures \cite{Rossinelli_etal_2015}.

The main idea of these methods is the use of a multiresolution data representation.
The decay of the detail coefficients, which describe the difference between two subsequent resolutions, yields information on local regularity of the solution.
Thus the truncation error can be estimated and grids can be coarsened in regions where this error is small and the solution is smooth.
Thresholding the multiresolution representation allows to introduce easily such adaptive grids where only significant coefficients are
retained. Hence uniform grid computations can be accelerated considerably as the number of points can be significantly reduced,
while controlling the accuracy of the discretization. Memory requirements could also be reduced if dynamic data structures are used.
Possible approaches to data structure management include the use of space filling curves \cite{Dahmen_etal_2013} or hash tables \cite{Brun_etal_2012}.
Our method is related to earlier work by Roussel and Schneider \cite{Roussel_Schneider_2010} and it uses tree data structures.

Adaptive tree codes have been used for particle methods, e.g., for computing vortex sheet roll-up \cite{Lindsay_Krasny_2001}.
Quad/octree based adaptive solvers for the time-dependent incompressible
Euler equations, even coupled with volume-of-fluid techniques, have been
proposed in \cite{Popinet_2003,Popinet_2009}.
In the context of level set methods, quadtree and octree data structures have been used for adaptive (non-graded) Cartesian grids \cite{Min_Gibou_2007}. Even water and smoke simulations have been performed exploiting octree data structures and mesh refinement \cite{Losasso_etal_2004}.
Reviews on different multiresolution methods can be found,
e.g., in the books of Cohen \cite{Cohen_2000} and M\"{u}ller \cite{Muller_2003} or in the overview article
by Domingues et al. \cite{Domingues_etal_2011}.

The paper is organized as follows. Section~\ref{sec:levelset_method} describes the gradient-augmented level set method.
Section~\ref{sec:multiresolution} briefly presents the multiresolution analysis. Section~\ref{sec:tree} introduces the tree data structure.
Section~\ref{sec:time_stepping} discusses adaptive time stepping techniques.
Section~\ref{sec:algorithm} summarizes the algorithm.
Numerical validation and performance tests are presented in sections~\ref{sec:multi_validation} through \ref{sec:performance}.
Section~\ref{sec:conclusions} draws some conclusions and presents possible perspectives for future work.

\section{Problem definition and description of the method}\label{sec:method}

\subsection{Gradient-augmented level-set method for advection problems}\label{sec:levelset_method}

The gradient-augmented level set method \cite{Seibold_etal_2012,Nave_etal_2010,Chidyagwal_etal_2012} is an efficient tool for numerical solution of advection problems.
In this work, we consider the linear advection equation
\begin{equation}
u_t + \bm{a} \cdot \nabla u = 0, \quad \mathrm{for}~ t>0,~ x \in \Omega \subset \mathbb{R}^2,
\label{eq:levelset}
\end{equation}
with suitable initial and boundary conditions. In (\ref{eq:levelset}), $u(\bm{x},t)$ is a scalar valued function,
$\bm{a}(\bm{x},t)=\left(a_1,a_2\right)$ is a velocity field, $\bm{x}=(x_1,x_2)$ is the position vector and $t$ is time.
In the present paper we only discuss two-dimensional problems, but it is straightforward to generalize the numerical method to three dimensions.

In practice, the method can be used to solve problems that have nonsmooth solutions.
In the context of closely related Hermite methods, this class of problems was treated in \cite{Appelo_Hagstrom_2011}.
However, here we assume that $u(\bm{x},t)$ is smooth, to avoid additional complications.

\begin{figure}[htb]
\centering
\includegraphics[scale=0.8]{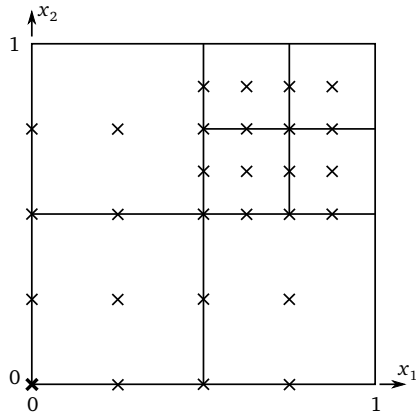}
\caption{Example of a discretization grid. Markers $\times$ show grid points $\bm{x}_j$.}\label{fig:drawing_grid}
\end{figure}

The numerical method consists in solving an augmented system of equations.
In addition to the level-set function $u$, its partial derivatives $u_{x_1}$, $u_{x_2}$ and $u_{x_1 x_2}$ are also evolved.
The corresponding evolution equations are obtained by differentiating (\ref{eq:levelset}).
All quantities are stored at discrete adaptive grid points $\{\bm{x}_j\}_{j=1,J}$.
An example of discretization grid is shown in figure~\ref{fig:drawing_grid}.
In this paper, we assume that the computational domain $\Omega$ in space is a unit square. The results may be easily rescaled to smaller or larger domains.
Adaptivity and remeshing techniques are discussed in sections~\ref{sec:multiresolution} and \ref{sec:tree}.
However, in the current section, it is assumed that the discretization grid $\{\bm{x}_j\}_{j=1,J}$ is known.

We use a semi-Lagrangian approach \cite{Seibold_etal_2012}.
Evolution of $u$ and its derivatives along the characteristic lines is described by a system of ODEs,
\begin{equation}
\frac{\mathrm{d}\bm{X}}{\mathrm{d}t} = \bm{a}(\bm{X},t)
\label{eq:characteristics}
\end{equation}
with appropriate initial conditions. At each time step, we consider points $\bm{x}^\textrm{sw}_j=\bm{x}_j + \eta \bm{d}^\textrm{sw}$,
$\bm{x}^\textrm{se}_j=\bm{x}_j + \eta \bm{d}^\textrm{se}$,
$\bm{x}^\textrm{nw}_j=\bm{x}_j + \eta \bm{d}^\textrm{nw}$ and
$\bm{x}^\textrm{ne}_j=\bm{x}_j + \eta \bm{d}^\textrm{ne}$,
where
$\bm{d}^\textrm{sw} = (-1,-1)$, $\bm{d}^\textrm{se} = (1,-1)$, $\bm{d}^\textrm{nw} = (-1,1)$, $\bm{d}^\textrm{ne} = (1,1)$,
and $\eta$ is a small number compared to the minimum grid step size. We set $\eta=10^{-7}$ in the examples shown in this paper, unless stated otherwise.
The effect of varying $\eta$ is discussed in section~\ref{sec:conv_adaptive}.
Characteristics that go through grid points $\bm{x}^\textrm{sw}_j$, $\bm{x}^\textrm{se}_j$, $\bm{x}^\textrm{nw}_j$ and $\bm{x}^\textrm{ne}_j$
are traced backwards in time from $t_{n+1}$ to $t_n=t_{n+1}-\Delta t$.
For every $j$, the solution $\bm{X}_j^{\textrm{ft sw}}=\bm{X}(t_n)$ of the final value problem
\begin{equation}
\left\{\begin{array}{l}
\displaystyle \frac{\textrm{d}}{\textrm{d}\tau} \bm{X}(\tau) = \bm{a}(\bm{X}(\tau),\tau) \\
\displaystyle \bm{X}(t_{n+1}) = \bm{x}_j^\textrm{sw}
\end{array}
\right.
\label{eq:num_char}
\end{equation}
is obtained approximately using one step of a Runge--Kutta scheme, as will be discussed in greater detail in section~\ref{sec:time_stepping}.
Points $\bm{X}_j^{\textrm{ft se}}$, $\bm{X}_j^{\textrm{ft nw}}$ and  $\bm{X}_j^{\textrm{ft ne}}$ are found similarly.
Their average is then calculated,
\begin{equation}
 \bm{X}_j^{\textrm{ft}} = \frac{1}{4} \left( \bm{X}_j^{\textrm{ft sw}} + \bm{X}_j^{\textrm{ft se}} + \bm{X}_j^{\textrm{ft nw}} + \bm{X}_j^{\textrm{ft ne}} \right).
\label{eq:ft_ave_grad}
\end{equation}
Values of $u(\bm{X}_j^{\textrm{ft sw}},t_n)$, $u(\bm{X}_j^{\textrm{ft se}},t_n)$, $u(\bm{X}_j^{\textrm{ft nw}},t_n)$ and $u(\bm{X}_j^{\textrm{ft ne}},t_n)$
are calculated by Hermite interpolation using the known grid-point values (see appendix~\ref{sec:appendix}).
It is important that the same interpolant is used for calculating all of these four points, as was pointed out in \cite{Seibold_etal_2012}.
We use the interpolant defined by the cell that contains $\bm{X}_j^{\textrm{ft}}$.

Then the grid-point values of $u$ at time $t_{n+1}$ are obtained by averaging, and the corresponding derivatives are calculated using second-order finite-difference approximations \cite{Seibold_etal_2012},
\begin{equation}
\begin{array}{l}
\displaystyle u(\bm{x}_j,t_{n+1}) = \frac{1}{4} \left( u(\bm{X}_j^{\textrm{ft sw}},t_n) + u(\bm{X}_j^{\textrm{ft se}},t_n) + u(\bm{X}_j^{\textrm{ft nw}},t_n) + u(\bm{X}_j^{\textrm{ft ne}},t_n) \right), \\
\displaystyle u_{x_1}(\bm{x}_j,t_{n+1}) = \frac{1}{4 \eta} \left( - u(\bm{X}_j^{\textrm{ft sw}},t_n) + u(\bm{X}_j^{\textrm{ft se}},t_n) - u(\bm{X}_j^{\textrm{ft nw}},t_n) + u(\bm{X}_j^{\textrm{ft ne}},t_n) \right),  \\
\displaystyle u_{x_2}(\bm{x}_j,t_{n+1}) = \frac{1}{4 \eta} \left( - u(\bm{X}_j^{\textrm{ft sw}},t_n) - u(\bm{X}_j^{\textrm{ft se}},t_n) + u(\bm{X}_j^{\textrm{ft nw}},t_n) + u(\bm{X}_j^{\textrm{ft ne}},t_n) \right),  \\
\displaystyle u_{x_1 x_2}(\bm{x}_j,t_{n+1}) = \frac{1}{4 \eta^2} \left( u(\bm{X}_j^{\textrm{ft sw}},t_n) - u(\bm{X}_j^{\textrm{ft se}},t_n) - u(\bm{X}_j^{\textrm{ft nw}},t_n) + u(\bm{X}_j^{\textrm{ft ne}},t_n) \right).  \\
\end{array}
\label{eq:update_rule}
\end{equation}

We assume the initial condition being prescribed analytically and being sufficiently regular. Therefore, we have access to the exact values of
$u(\bm{x}_j,t_0)$, $u_{x_1}(\bm{x}_j,t_0)$, $u_{x_2}(\bm{x}_j,t_0)$ and $u_{x_1 x_2}(\bm{x}_j,t_0)$
required for startup.

In this work we only consider $u$ and $\bm{a}$ periodic in space.
Implementation of Dirichlet or Neumann boundary conditions is less straightforward, but possible, as discussed in \cite{Nave_etal_2010}.

\subsection{Multiresolution analysis for error estimate}\label{sec:multiresolution}

To obtain an error estimate required for mesh adaptation, we use discrete multiresolution analysis.
Interpolatory multiresolution analysis based on Hermite interpolation was studied by Warming and Beam \cite{Warming_Beam_2000}.
It is consistent with our numerical method as the gradient information
is available. Hence, advecting the function values and its derivatives requires error control for both quantities.
Fortunately, the computational overhead due to the error estimate is small, since the mid-point interpolation formulae
are much simpler than interpolation at an arbitrary point, see appendix~\ref{sec:appendix}.

\begin{figure}[htb]
\centering
\includegraphics[scale=0.8]{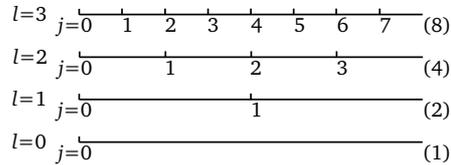}
\caption{One-dimensional nested grids. In this example, $m=0$ and $M=3$.
Note that, by periodicity, the right end point of the interval is
identical to the left end point.}\label{fig:drawing_1d}
\end{figure}

For introduction, let us first consider a one-dimensional multiresolution transform
of data sampled on a uniform grid consisting of $2^M$ points. Let $\{ x_j^l \}_{j=0,2^l}$ be a nested sequence of uniform dyadic grids on the unit interval $[0,1]$, such that
\begin{equation}
x_j^l = j h_l, \quad h_l = 1/2^l
\label{eq:nested_grids}
\end{equation}
and $l=m,m+1,...,M$, where $m$ is the coarsest and $M$ is the finest level index.
An example is shown in figure~\ref{fig:drawing_1d}.
It follows that the grid at level $l-1$ is formed from the grid at level $l$
by removing grid points with odd indices:
\begin{equation}
x_j^{l-1} = x_{2j}^l, \quad j = 0, 1, 2, ..., 2^{l-1}.
\end{equation}

Let
\begin{equation}
u_j^M = u( x_j^M ), \quad {u'}_j^M = u'( x_j^M ), \quad j=0,...,2^M,
\label{eq:mra_point_values}
\end{equation}
be the finest-grid point values of a scalar function and its derivative.
It is convenient to scale the derivative by defining
\begin{equation}
v_j^M = h_M {u'}_j^M.
\end{equation}
Note that, by periodicity, $u_{2^l}^l = u_0^l$ and $v_{2^l}^l = v_0^l$ for all $l$.
Scaling is also required for stability of the multiresolution transform, when $M-m$ is large, but in our present work this is not a constraint.

From the point values, the multiresolution transform calculates
coarsest-level values of $u$ and $v$ and their details at all levels,
\begin{equation}
\begin{array}{l}
\displaystyle \left\{ u_0^m,...,u_{2^m-1}^m; r_0^m,...,r_{2^m-1}^m; r_0^{m+1},...,r_{2^{m+1}-1}^{m+1}; r_0^{M-1},...,r_{2^{M-1}-1}^{M-1} \right\}, \\
\displaystyle \left\{ v_0^m,...,v_{2^m-1}^m; s_0^m,...,s_{2^m-1}^m; s_0^{m+1},...,s_{2^{m+1}-1}^{m+1}; s_0^{M-1},...,s_{2^{M-1}-1}^{M-1} \right\}.
\end{array}
\end{equation}
It is straightforward to project point values and derivative values at
even-numbered grid points from level $l$ to $l-1$,
\begin{equation}
u_j^{l-1} = u_{2j}^l, \quad v_j^{l-1} = 2v_{2j}^l, \quad j=0,2,...,2^{l-1}-1.
\end{equation}
The details at level $l-1$ are calculated as the difference between the exact and the interpolated values at odd points at level $l$,
\begin{equation}
r_j^{l-1} = u_{2j+1}^l - \tilde{u}_{2j+1}^l, \quad s_j^{l-1} = v_{2j+1}^l - \tilde{v}_{2j+1}^l,
\end{equation}
where $\tilde{u}_{2j+1}^l$ and $\tilde{v}_{2j+1}^l$ are calculated by Hermite interpolation,
\begin{equation}
\begin{array}{l}
\displaystyle \tilde{u}_{2j+1}^l = \frac{1}{2} \left( u_{2j}^l + u_{2j+2}^l \right) + \frac{1}{4} \left( v_{2j}^l - v_{2j+2}^l \right), \\
\displaystyle \tilde{v}_{2j+1}^l = - \frac{3}{4} \left( u_{2j}^l - u_{2j+2}^l \right) - \frac{1}{4} \left( v_{2j}^l + v_{2j+2}^l \right).
\end{array}
\end{equation}

The advantage of the multiresolution representation is that many details are small or even zero in regions in which the function $u$ is smooth. Thus high data compression ratios can be obtained for functions with inhomogeneous regularity, i.e., their Besov regularity is larger than their Sobolev regularity \cite{DeVore_1998}.

\begin{figure}[htb]
\centering
\includegraphics[scale=0.8]{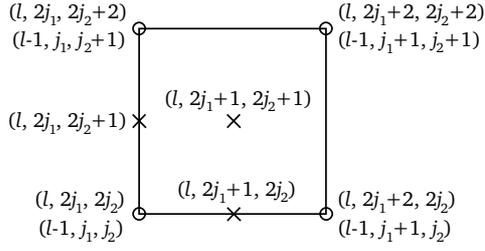}
\caption{Discretization grid cell at $(l-1,j_1,j_2)$.
Markers $\times$ denote the 3 points where values of the function $u$ and its derivatives are stored and residuals are computed.
Markers $\circ$ denote the corner points that are used for interpolation.}\label{fig:drawing_cell}
\end{figure}

In \cite{Warming_Beam_2000}, only a one-dimensional multiresolution transform was considered.
We now discuss the two-dimensional case.
We use Cartesian coordinates $\bm{x}=(x_1,x_2)$.
Let $\{ \bm{x}_{j_1,j_2}^l \}_{j_1=0,2^l; j_2=0,2^l}$, be a uniform dyadic grid on $[0,1]\times[0,1]$.
This grid consists of points
\begin{equation}
\bm{x}_{j_1,j_2}^l = ( j_1 h_l, j_2 h_l ), \quad h_l = 1/2^l.
\label{eq:nested_grids_2d}
\end{equation}
We consider a nested sequence of such grids that correspond to levels $l=m,m+1,...,M$.
It is assumed that values of the function and its scaled partial derivatives are given on the finest grid,
\begin{equation}
\begin{array}{l}
\displaystyle (u_0)_{j_1,j_2}^M = u( \bm{x}_{j_1,j_2}^M ), \\
\displaystyle (u_1)_{j_1,j_2}^M = h_M \frac{\partial u}{\partial x_1}( \bm{x}_{j_1,j_2}^M ), \quad (u_2)_{j_1,j_2}^M = h_M \frac{\partial u}{\partial x_2}( \bm{x}_{j_1,j_2}^M ), \\
\displaystyle (u_3)_{j_1,j_2}^M = h_M^2 \frac{\partial^2 u}{\partial x_1 \partial x_2}( \bm{x}_{j_1,j_2}^M ),
\end{array}
\label{eq:mra_point_values_2d}
\end{equation}
where $j_1=0,...,2^M$ and $j_2=0,...,2^M$.
Their values are projected to coarser levels and, at every level, horizontal, vertical and diagonal details are computed.
This requires interpolation at points $(l,2j_1+1,2j_2)$, $(l,2j_1,2j_2+1)$ and $(l,2j_1+1,2j_2+1)$, respectively,
using the coarser-grid values at points $(l-1,j_1,j_2)$, $(l-1,j_1+1,j_2)$, $(l-1,j_1,j_2+1)$ and $(l-1,j_1+1,j_2+1)$, as explained in figure~\ref{fig:drawing_cell}.

Thus we obtain an algorithm for a two-dimensional multiresolution decomposition,
\begin{equation}
\begin{array}{l}
\textbf{for } l = M,M-1,...,m+1 \\
\quad \textbf{for } j_1 = 0,1,...,2^{l-1}-1 \\
\quad\quad \textbf{for } j_2 = 0,1,...,2^{l-1}-1 \\
\quad\quad\quad \textbf{for } \iota = 0,1,2,3 \\
\quad\quad\quad\quad (u_\iota)_{j_1,j_2}^{l-1} = \alpha_\iota (u_\iota)_{2j_1,2j_2}^l \\
\quad\quad\quad\quad (r_\iota^1)_{j_1,j_2}^{l-1} = (u_\iota)_{2j_1+1,2j_2}^l - (\tilde{u}_\iota)_{2j_1+1,2j_2}^l \\
\quad\quad\quad\quad (r_\iota^2)_{j_1,j_2}^{l-1} = (u_\iota)_{2j_1,2j_2+1}^l - (\tilde{u}_\iota)_{2j_1,2j_2+1}^l \\
\quad\quad\quad\quad (r_\iota^3)_{j_1,j_2}^{l-1} = (u_\iota)_{2j_1+1,2j_2+1}^l - (\tilde{u}_\iota)_{2j_1+1,2j_2+1}^l \\
\quad\quad\quad \textbf{end} \\
\quad\quad \textbf{end} \\
\quad \textbf{end} \\
\textbf{end}
\end{array}
\label{eq:decomposition}
\end{equation}
and for reconstruction,
\begin{equation}
\begin{array}{l}
\textbf{for } l = m+1,m+2,...,M \\
\quad \textbf{for } j_1 = 0,1,...,2^{l-1}-1 \\
\quad\quad \textbf{for } j_2 = 0,1,...,2^{l-1}-1 \\
\quad\quad\quad \textbf{for } \iota = 0,1,2,3 \\
\quad\quad\quad\quad (u_\iota)_{2j_1,2j_2}^l = (u_\iota)_{j_1,j_2}^{l-1} / \alpha_\iota \\
\quad\quad\quad\quad (u_\iota)_{2j_1+1,2j_2}^l = (\tilde{u}_\iota)_{2j_1+1,2j_2}^l + (r_\iota^1)_{j_1,j_2}^{l-1} \\
\quad\quad\quad\quad (u_\iota)_{2j_1,2j_2+1}^l = (\tilde{u}_\iota)_{2j_1,2j_2+1}^l + (r_\iota^2)_{j_1,j_2}^{l-1} \\
\quad\quad\quad\quad (u_\iota)_{2j_1+1,2j_2+1}^l = (\tilde{u}_\iota)_{2j_1+1,2j_2+1}^l + (r_\iota^3)_{j_1,j_2}^{l-1} \\
\quad\quad\quad \textbf{end} \\
\quad\quad \textbf{end} \\
\quad \textbf{end} \\
\textbf{end}
\end{array}
\label{eq:reconstruction}
\end{equation}
In the above, $\alpha_0 = 1$, $\alpha_1 = \alpha_2 = 2$ and $\alpha_3 = 4$.
The two-dimensional interpolation formulae for $\tilde{u}_\iota$ are given in appendix~\ref{sec:appendix}.
The computational complexity of the above multiresolution transform and its inverse is linear,
since each of the details is only accessed once, and
the number of details scales as the number of finest-level grid points.

In the grid adaptation process, which is part of the numerical method for the advection equation proposed in this work,
only the multiresolution decomposition is used. The reconstruction procedure is required to obtain the solution on a regular grid to perform the error analysis,
and thus we use the algorithm (\ref{eq:reconstruction}) in our validation tests in section~\ref{sec:multi_validation}.

The outer loop in the multiresolution transform (\ref{eq:decomposition})
is defined from the finest possible level $M$ down to one level above the coarsest, $m+1$.
This algorithm assumes that grid-point data are available at the finest level.
However, for the nonuniform adaptive grid described in the following sections, the finest level
varies depending on the position $(j_1,j_2)$. Therefore, in that case, the decomposition
starts from the local finest level at the given position. Also, when multiresolution is used for grid adaptation,
it is unnecessary to compute the decomposition for all levels down to $m+1$.
It is only computed for two levels downwards. These technicalities are described in section~\ref{sec:algorithm}.

The magnitude of details $(r_\iota^i)_{j_1,j_2}^l$ decreases with level $l$,
with a rate that depends on the regularity of the function $u$ and on the accuracy of interpolation.
This property is used for data compression.
We define a truncation operator,
\begin{equation}
(\hat{r}_\iota^i)_{j_1,j_2}^l =
\left\{
\begin{array}{lll}
(r_\iota^i)_{j_1,j_2}^l & \textrm{if } |(r_\iota^i)_{j_1,j_2}^l| > \varepsilon_l & \textrm{for any } i, \iota \\
0                       & \textrm{if } |(r_\iota^i)_{j_1,j_2}^l| \le \varepsilon_l & \textrm{for every } i, \iota
\end{array}
\right.
\label{eq:truncation}
\end{equation}
where $0 \le \iota \le 3$ and $1 \le i \le 3$ and $\varepsilon_l$ is a threshold defined below.
As one can see from (\ref{eq:mra_point_values_2d}) and (\ref{eq:decomposition}), the index $\iota$
denotes the quantity for which the detail is computed (0: function value; 1: first derivative with respect to $x_1$;
2: first derivative with respect to $x_2$; 3: cross derivative with respect to $x_1$ and $x_2$).
The index $i$ denotes the direction of the detail (1: horizontal; 2: vertical; 3: diagonal),
depending on which of the three points indicated by the ``$\times$" marker in figure~\ref{fig:drawing_cell} is taken to compute the detail.
In other words, in the context of the tree data structure discussed in the following section, there are 12 details per cell, and a cell with its three inner points can only be discarded if all of the details are less than $\varepsilon_l$ in magnitude.

In practice, only non-zero values of $(\hat{r}_\iota^i)_{j_1,j_2}^l$ are stored and
used in the reconstruction algorithm (\ref{eq:reconstruction}) instead of $(r_\iota^i)_{j_1,j_2}^l$.
Thus, the obtained approximate values of $(\hat{u}_\iota^i)_{j_1,j_2}^M$ differ from the exact values $(u_\iota^i)_{j_1,j_2}^M$
and the error depends on the choice of $\varepsilon_l$.
In general, $\varepsilon_l$ may depend on $l$.
However, in our method the threshold is scale-independent,
\begin{equation}
\varepsilon_l = \varepsilon,
\label{eq:cutoff_scind}
\end{equation}
which is required to control the error in the $L^\infty$ norm.
Later on, we also relate the time discretization
error control to $\varepsilon$, and show some numerical evidence
that the local error of the method at each time step is indeed proportional to $\varepsilon$.

\subsection{Tree data structure}\label{sec:tree}

The performance of an adaptive numerical method strongly depends on the data structures, which are necessary for memory compression.
For two-dimensional multiresolution, it is natural to use quadtrees.
Even though the quadtree is a classical data structure,
the way it is associated with an adaptive grid depends on the underlying numerical method.
In this section we briefly describe its implementation for adaptive methods based on point-value multiresolution.

\begin{figure}[htb]
\centering
\includegraphics[scale=0.8]{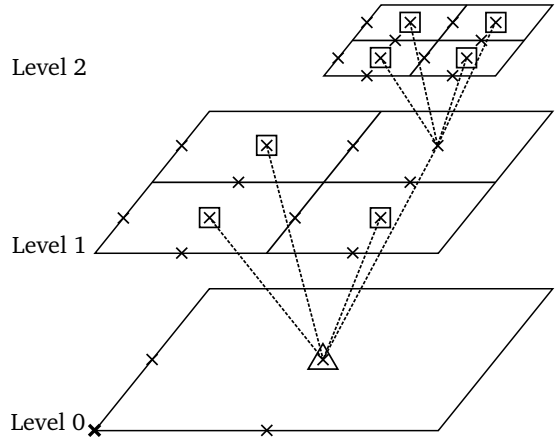}
\caption{Tree data structure. Every node of the tree corresponds to a Cartesian grid cell.
Squares indicate the leaves and a triangle indicates the root of the tree.
Markers $\times$ show grid points associated with the nodes of the tree (or cells).
The bold marker $\bm{\times}$ denotes the point at the origin that is not associated with the tree.}\label{fig:drawing_tree}
\end{figure}

We define a quadtree node to be a discretization grid cell, as shown in figure~\ref{fig:drawing_tree}.
A node in the tree is indexed by level $l$ and position $j_1$, $j_2$.
The node at level $0$ is called the root of the tree.
A node at level $l-1$ may have 4 child nodes at level $l$. This corresponds to a partition of a parent cell into 4 child cells.
In our implementation, a node either has four or no children.
A node that does not have children is referred to as a leaf. The discretization grid is formed by the leaves.
We construct graded trees such that the level difference between two neighbouring leaves is not greater than one. We use an algorithm described in \cite{Wang_2007}.

Vertices at cell corner points store values of $u$, $u_{x_1}$, $u_{x_2}$ and $u_{x_1 x_2}$. Thus a cell contains all information required for Hermite interpolation (see appendix~\ref{sec:appendix}).
Note that many cells corresponding to different nodes at different levels may point at the same vertex that they have in common.
An efficient algorithm is required to store and to access the grid point values.
For interpolating multiresolution with graded trees, there is a one-to-one correspondence between quadtree nodes (or cells) and grid points \cite{Harten_1996}.
A cell at $(l-1,j_1,j_2)$ has three points assigned to it, as shown in figure~\ref{fig:drawing_cell}.
Their Cartesian coordinates are $\left((2j_1+1) h_l, 2j_2 h_l\right)$, $\left(2j_1 h_l, (2j_2+1) h_l\right)$ and $\left((2j_1+1) h_l, (2j_2+1) h_l\right)$, where $h_l = 1/2^l$, $j_1,j_2 = 0,...,2^{l-1}-1$.
These points are also corners of cells at level $l$ or higher.
To access a grid point, an algorithm similar to a tree search is employed, that requires $\mathcal{O}(\log N)$ operations, where $N$ is the number of grid points.
We note a possibility of reducing the cost of access to point data down to $\mathcal{O}(1)$ by using other kinds of
data structures, such as space filling curves \cite{Dahmen_etal_2013} or hash tables \cite{Brun_etal_2012}.

The list of all grid points plus the point at the origin is denoted as $\mathcal{L}_{points}$.
If the tree is graded, and since the computational domain is a torus, the corner points of any cell coincide with 4 points from $\mathcal{L}_{points}$.
Gradedness is required for several and different reasons.
First, since we use nested grids obtained by cells with three interior points assigned to each cell, a non-graded tree can lead to a situation when there is no grid point at the corner of a cell.
Second, after coarsening, we only add one level of refinement. Non-graded trees would require adding as many levels of refinement as the difference in level between neighbouring cells.
Third, large differences in the size of neighbouring cells may have a negative effect on the stability of discretization schemes. However, for our scheme we do not have any clear evidence of this effect, and in the case of a one-dimensional advection equation the method is stable even when non-graded binary trees are employed.

\subsection{Adaptive time stepping}\label{sec:time_stepping}

The numerical method discussed in section~\ref{sec:levelset_method} does not have any stability restriction on the
time step size $\Delta t$. However, for a given discretization in space, there exists an optimal $\Delta t$
that minimizes the error. If $\Delta t$ is too large, the time discretization error becomes large.
If $\Delta t$ is too small, the error becomes large because of accumulation of space
discretization errors at every time step, such that the global error at the final time step grows in proportion to
the number of time steps. This optimality condition is satisfied if the time step varies in accordance
with the space grid size locally. Local time stepping is currently used in adaptive methods for PDEs (see, \eg, \cite{Domingues_etal_2008}),
but its implementation is not staightforward for higher order time discretization schemes and we consider it as a possible future work.

In the adaptive time-stepping algorithms presented in this paper,
the solution is evolved in time with the same $\Delta t$ for all grid points.
At each time iteration, after taking a step of size $\Delta t$, a new adapted value of time step size is calculated, which we denote $\Delta t^*$.
We use the Dormand--Prince Runge--Kutta 4(3) method \cite{Dormand_Prince_1978},
that gives a local error estimate required for time step size selection.
The third order error estimate of this scheme is more reliable
at moderate step sizes $\Delta t$, compared to the fifth order estimate of the Runge--Kutta--Fehlberg method (see \cite{Dormand_Prince_1978}).
When written backwards in time, its coefficients read
\begin{equation}
\begin{array}{lcl}
\bm{k}_1 & = & - \Delta t \bm{a}(\bm{x}_j^\textrm{dir} ,t_{n+1}) \\
\bm{k}_2 & = & - \Delta t \bm{a}(\bm{x}_j^\textrm{dir}+\frac{1}{2}\bm{k}_1 ,t_{n+1}-\frac{1}{2}\Delta t) \\
\bm{k}_3 & = & - \Delta t \bm{a}(\bm{x}_j^\textrm{dir}+\frac{1}{2}\bm{k}_2 ,t_{n+1}-\frac{1}{2}\Delta t) \\
\bm{k}_4 & = & - \Delta t \bm{a}(\bm{x}_j^\textrm{dir}+\bm{k}_3 ,t_{n+1}-\Delta t)
\end{array}
\label{eq:rk4_coefs}
\end{equation}
where $\bm{x}_j^\textrm{dir}$ are points of the finite-difference stencil introduced in section~\ref{sec:levelset_method}, subscript `dir' refers to a direction
of shift relative to the $j$-th grid point (`sw', `se', `nw' or `ne').
The coordinates of the corresponding Lagrangian point are calculated as
\begin{equation}
\bm{X}_j^{\textrm{ft dir}} = \bm{x}_j^\textrm{dir} + \frac{1}{6}\bm{k}_1 + \frac{1}{3}\bm{k}_2 + \frac{1}{3}\bm{k}_3 + \frac{1}{6}\bm{k}_4
\label{eq:rk4_point}
\end{equation}
with the local truncation error $\mathcal{O}(\Delta t^5)$.
This classical scheme is embedded in a five-step third-order formula that provides a reliable error estimate,
\begin{equation}
\bm{T}_j^{\textrm{dir}} = \lambda \left( \bm{k}_4 + \Delta t \bm{a}(\bm{X}_j^{\textrm{ft dir}} ,t_{n+1}-\Delta t) \right),
\label{eq:rk4_error}
\end{equation}
where $\lambda = 1/10$, as suggested by \cite{Dormand_Prince_1978}.
The error estimate is averaged in the four directions,
\begin{equation}
\bm{T}_j = \frac{1}{4} \left( \bm{T}_j^{\textrm{sw}} + \bm{T}_j^{\textrm{se}} + \bm{T}_j^{\textrm{nw}} + \bm{T}_j^{\textrm{ne}} \right),
\end{equation}
and used in a Taylor expansion of $u$ that provides a local truncation error estimate to the semi-Lagrangian part of the method,
\begin{equation}
E_j = u_{x_1}(\bm{x}_j,t_{n+1}) {T_1}_j + u_{x_2}(\bm{x}_j,t_{n+1}) {T_2}_j + u_{x_1 x_2}(\bm{x}_j,t_{n+1}) {T_1}_j {T_2}_j,
\label{eq:error_estimate}
\end{equation}
where the derivatives of $u$ at time $t$ are calculated according to (\ref{eq:update_rule}).
Then the error norm is calculated, $||E||_\infty = \max_{\bm{x}_j \in \mathcal{L}_{points}} |E_j|$, where $\mathcal{L}_{points}$
is a list of all grid points.

The new adapted time step size $\Delta t^*$ is
\begin{equation}
\Delta t^* = \min\left( 1~,~ \Delta t  \max\left(0.5~, \min\left( 2~,~ 0.75 \left( \frac{\varepsilon}{||E||_\infty} \right)^{\frac{1}{4}} \right) \right) \right).
\label{eq:new_time_step_size}
\end{equation}
It is limited above by 1, its growth rate is limited by 2 and its decrease rate is limited by 0.5.
The safety coefficient 0.75 helps to reduce the number of rejected time steps, as defined in the next paragraph.

If $||E||_\infty \le \varepsilon$, where $\varepsilon$ is the threshold, the values $u(\bm{x}_j,t)$, $u_{x_1}(\bm{x}_j,t)$, $u_{x_2}(\bm{x}_j,t)$ and $u_{x_1 x_2}(\bm{x}_j,t)$
are assumed being sufficiently accurate and the computation proceeds to the next time iteration with
new step size $\Delta t = \Delta t^*$.
Otherwise, the time step is rejected, and these quantities are recalculated using (\ref{eq:rk4_coefs}), (\ref{eq:rk4_point}) and (\ref{eq:update_rule}) with step size $\Delta t^*$ instead of $\Delta t$.
As a result, the
error estimate (\ref{eq:error_estimate}) and the new time step size (\ref{eq:new_time_step_size}) are computed again.
The process is repeated until convergence.
In general, we observe convergence within a few iterations.

The initial choice of $\Delta t$ at $t=0$ is important.
It should be sufficiently small to ensure a good accuracy of error estimates based on Taylor series expansions.
Thus, after building the initial grid and initializing the grid point values of the solution at $t=0$,
for each grid point we
determine the nearest cell size $h_j$ (of the four nearest cells, take the first found by tree search) and the velocity $\bm{a}_j=\bm{a}(\bm{x}_j,t)$.
Then we compute an approximation to the locally optimal time step size,
\begin{equation}
\Delta t_j = \frac{2 h_j}{\sqrt{|\bm{a}_j|^2+1}}.
\end{equation}
When $|\bm{a}_j|$ is large, this approximation coincides with the local CFL condition with the Courant number equal to 2. When $|\bm{a}_j|$ is small, $\Delta t_j$ is limited by twice the space step $h_j$.
The initial time step size $\Delta t|_{t=0}$ is the minimum of these local estimates,
\begin{equation}
\Delta t|_{t=0} = \min_{\bm{x}_j \in \mathcal{L}_{points}} \Delta t_j.
\label{eq:dt_init}
\end{equation}

\subsection{Algorithm and implementation aspects}\label{sec:algorithm}

The implementation of the method is, in some aspects, similar to earlier work by Roussel and Schneider \cite{Roussel_Schneider_2010}.
In the present work, we also use tree data structures. However, the present numerical method operates on point values rather than cell averages.
Since the numerical approximation of the derivatives is based on Hermite interpolation, the same interpolant is used for the error estimate.
This naturally led us to the point-value vector multiresolution \cite{Warming_Beam_2000}.

The algorithm consists of a startup phase followed by time stepping.

\begin{enumerate}
\item \emph{Startup.}
\begin{enumerate}

\item Set the threshold $\varepsilon$, domain size $L$ and time span $T$.

\item Set the coarsest and the finest possible discretization levels, $m$ and $M$, respectively.

\item Pick a value $l_{init} \in [m,M]$. Construct a tree structure
with $l_{init}$ levels that corresponds to a uniform grid.
$l_{init}$ is the starting level for multiresolution analysis of the initial condition. Note that it may have to be greater than $m$.
The initial uniform grid must be fine enough to be sensitive to the small-scale features of the initial condition.
If both $l_{init}$ and $l_{init}+1$ are too coarse, the analysis will stop before capturing those small scales.

\item \label{item:listpoints} Create a list of grid points $\mathcal{L}_{points}$.

\item Evaluate the initial condition at each point in $\mathcal{L}_{points}$.
It is assumed that the initial condition for $u$ and its derivatives can be evaluated at any level with machine precision.
For example, it is given analytically.

\item \label{item:remesh} Remesh as described below.

\item Repeat steps (\ref{item:listpoints})-(\ref{item:remesh}) $M-l_{init}$ times.

\item Set time $t=0$.

\item Initialize time step size $\Delta t$ (\ref{eq:dt_init}).

\end{enumerate}

\item \emph{Time steps.}
\begin{enumerate}

\item \label{item:listpoints_timestep} Create a list of grid points $\mathcal{L}_{points}$.

\item \label{item:adjust_timestep} If $t + \Delta t > T$, adjust the time step size to $\Delta t = T-t$. Similar adjustment is made if it is required to evaluate the solution at a given time $t_{output}$.

\item \label{item:forall_points} For all points $\bm{x}_j \in \mathcal{L}_{points}$,
  \begin{itemize}
  \item compute $\bm{x}_j^{\textrm{sw}}$, $\bm{x}_j^{\textrm{se}}$, $\bm{x}_j^{\textrm{nw}}$ and $\bm{x}_j^{\textrm{ne}}$;
  \item using Runge-Kutta integration (\ref{eq:rk4_point}), compute $\bm{X}_j^{\textrm{ft sw}}$, $\bm{X}_j^{\textrm{ft se}}$, $\bm{X}_j^{\textrm{ft nw}}$ and $\bm{X}_j^{\textrm{ft ne}}$;
  \item estimate the local truncation error of time integration using (\ref{eq:rk4_error}): $\bm{T}_j^{\textrm{sw}}$, $\bm{T}_j^{\textrm{se}}$, $\bm{T}_j^{\textrm{nw}}$ and $\bm{T}_j^{\textrm{ne}}$;
  \item determine $u(\bm{X}_j^{\textrm{ft sw}},t_n)$, $u(\bm{X}_j^{\textrm{ft se}},t_n)$, $u(\bm{X}_j^{\textrm{ft nw}},t_n)$ and $u(\bm{X}_j^{\textrm{ft ne}},t_n)$ by Hermite interpolation;
  \item compute $u(\bm{x}_j,t+\Delta t)$, $u_{x_1}(\bm{x}_j,t+\Delta t)$, $u_{x_2}(\bm{x}_j,t+\Delta t)$ and $u_{x_1 x_2}(\bm{x}_j,t+\Delta t)$ from (\ref{eq:update_rule}) and store them in buffer variables;
  \item calculate error estimate $E_j$ (\ref{eq:error_estimate}).
  \end{itemize}

\item Compute $||E||_\infty$ using the local values $E_j$.

\item Calculate the new time step size $\Delta t^*$ given by (\ref{eq:new_time_step_size}).

\item If $||E||_\infty > \varepsilon$, assign $\Delta t = \Delta t^*$ and go to step (\ref{item:adjust_timestep}).

\item Increment time $t$ by $\Delta t$.

\item Assign $\Delta t = \Delta t^*$.

\item For all points $\bm{x}_j \in \mathcal{L}_{points}$, update the values of $u(\bm{x}_j,t)$, $u_{x_1}(\bm{x}_j,t)$, $u_{x_2}(\bm{x}_j,t)$ and $u_{x_1 x_2}(\bm{x}_j,t)$ using the values stored in the buffer variables.

\item \label{item:remesh_timestep} Remesh.

\item Repeat steps (\ref{item:listpoints_timestep})-(\ref{item:remesh_timestep}) until time $t$ reaches the final value $T$.

\end{enumerate}

\end{enumerate}

The \emph{remeshing} procedure is the same during the initial grid generation and during time iterations. It consists of the following steps.
\begin{enumerate}

\item \label{item:listleaves} Create a list of leaves of the tree structure, $\mathcal{L}_{leaves}$.
This is also a list of grid cells.
Note that the list $\mathcal{L}_{points}$ contains corner points of the cells in $\mathcal{L}_{leaves}$ as well
as their interior points that are used for computation of the residuals (see figure~\ref{fig:drawing_grid}).

\item For each cell in $\mathcal{L}_{leaves}$, use Hermite interpolation
to estimate the values of $u$ and its derivatives at the 3 interior points.
Compute the residuals, similarly to (\ref{eq:decomposition}).

\item Use the truncation formula (\ref{eq:truncation}) to determine which details can be discarded.
Mark the corresponding cells for coarsening.
Note that, in our implementation, coarsening implies removal of all four child cells of a parent cell.
Therefore coarsening is only allowed if all four are marked.

\item Ensure that the tree nodes are marked such that, after coarsening, the tree is graded.

\item Ensure that all cells at levels $l \le m$ are not marked and the cells at level $l=M$ are all marked for removal.

\item \label{item:coarsening} Coarsening: remove marked cells (\ie, marked nodes of the tree data structure).

\item Repeat steps (\ref{item:listleaves})-(\ref{item:coarsening}) once again.

\item Update list of leaves $\mathcal{L}_{leaves}$.

\item Refinement: split every leaf cell into four.
The uniform refinement method is, actually, the reason why two coarsening iterations are required.

\item Update list of leaves $\mathcal{L}_{leaves}$.

\item \label{item:} Assign new values to the interior points of all cells in $\mathcal{L}_{leaves}$ using Hermite interpolation of the corner-point values.
Note that interpolation inside the leaf cells does not result in any undesirable loss of accuracy because the corresponding details are less than $\varepsilon$, as ensured by the thresholding rule.
When handling the elements of $\mathcal{L}_{leaves}$, begin from the coarsest-level entries and end at the finest level,
because interpolation at finer levels uses point values at coarser levels.

\end{enumerate}

\section{Numerical results}\label{sec:numerics}

\subsection{Validation tests of the multiresolution transform}\label{sec:multi_validation}

\begin{figure}[htb]
\centering
\includegraphics[scale=0.8]{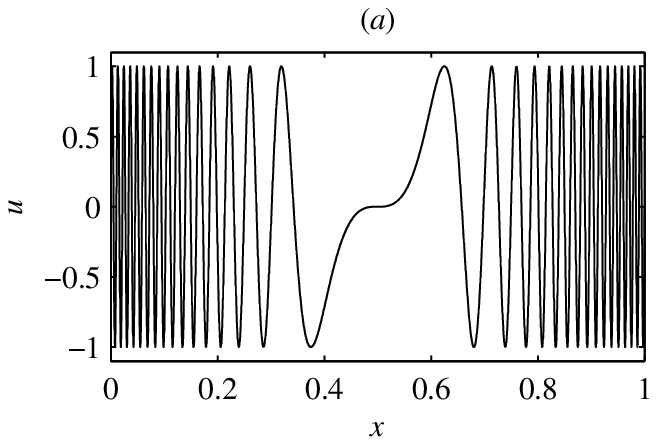}
\quad
\includegraphics[scale=0.8]{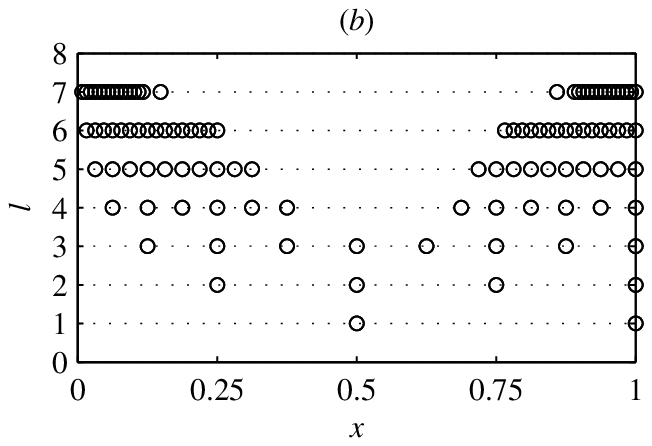}
\caption{(\textit{a}) Plot of a periodic chirp (\ref{eq:chirp}) with $\alpha=256$ and (\textit{b}) a diagram of its non-zero details after thresholding (\ref{eq:cutoff_scind}) with $\varepsilon=0.0664$.}\label{fig:chirp1d_a256}
\end{figure}

Let us consider a periodic one-dimensional chirp,
\begin{equation}
u = \sin\left( \alpha \pi (x-1/2)^3 \right), \quad 0 \le x \le 1,
\label{eq:chirp}
\end{equation}
where $\alpha$ is a parameter.
It is plotted in figure~\ref{fig:chirp1d_a256}(\textit{a}) for $\alpha=256$.

The function $u$ and its derivatives can be approximated with the desired accuracy, using only a finite number of details (\ie, grid points of an adaptive grid).
We remind that the multiresolution transform is based on the interpolation formulae that have
truncation error of order $\mathcal{O}(h^4)$, where $h=2^{-l}$.
Therefore, since the function (\ref{eq:chirp}) is smooth,
the global rate of decay of its details is $\mathcal{O}(2^{-4l})$.
Again, we remark that if the local regularity (measured in Besov spaces) of the function is larger than the global one (Sobolev) the detail coefficients even enjoy faster decay (see \cite{DeVore_1998}).
Here we compare the effect of nonlinear filtering, that is, discarding all details of magnitude less than $\varepsilon$,
and of linear filtering, that is, discarding all details at levels greater than $l_{max}$.
In the present case, more details are discarded with the nonlinear threshold, for a given accuracy in the $L^\infty$ norm.
This is illustrated in figure~\ref{fig:chirp1d_a256}(\textit{b}).
Open circles in the diagram display level $l$ and coordinate $x = (j+1)/ 2^l$ of the retained details.
The threshold $\varepsilon$ has been chosen such that the reconstructed function has the same error
as if all details at 7 levels were retained.
Details at $x \approx 0.5$ are smaller than those near $x=0$ and $x=1$ at the same level, hence they can be
discarded without loss of accuracy in the $L^\infty$ norm.

\begin{figure}[htbh]
\centering
\includegraphics[scale=0.8]{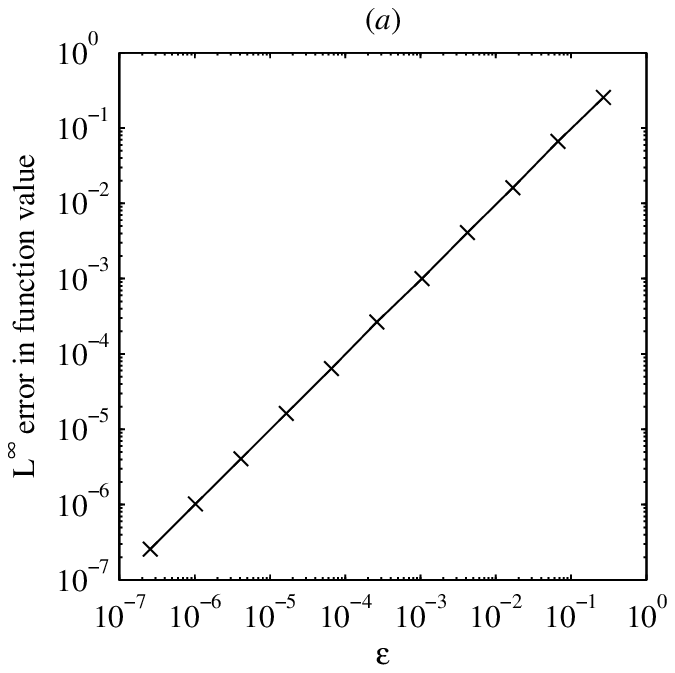}
\quad
\includegraphics[scale=0.8]{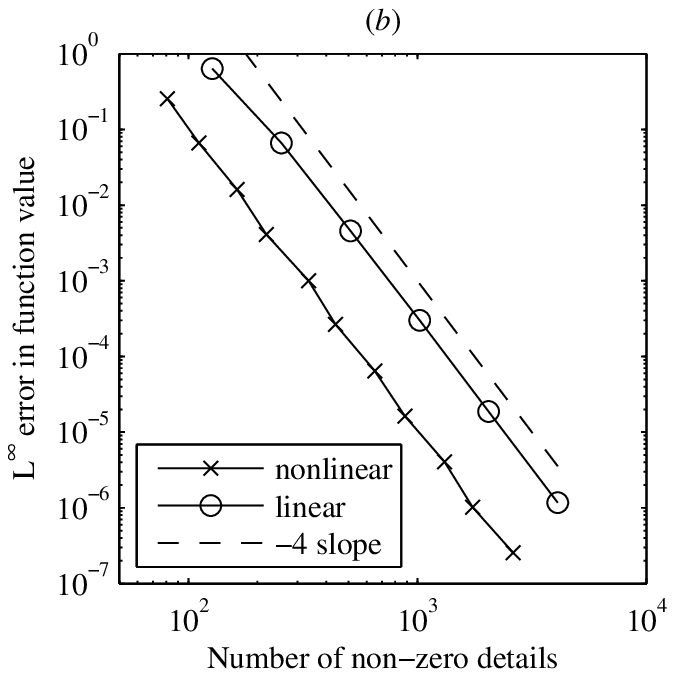}
\caption{(\textit{a}) Decay of $L^\infty$ error between the exact and reconstructed function values of a periodic chirp (\ref{eq:chirp}) with $\alpha=256$; (\textit{b}) decay of the $L^\infty$ error
versus the number of details retained by nonlinear (\ref{eq:cutoff_scind}) and linear filtering.}\label{fig:chirp1d_a256_errlinf}
\end{figure}

\begin{figure}[htbh]
\centering
\includegraphics[scale=0.8]{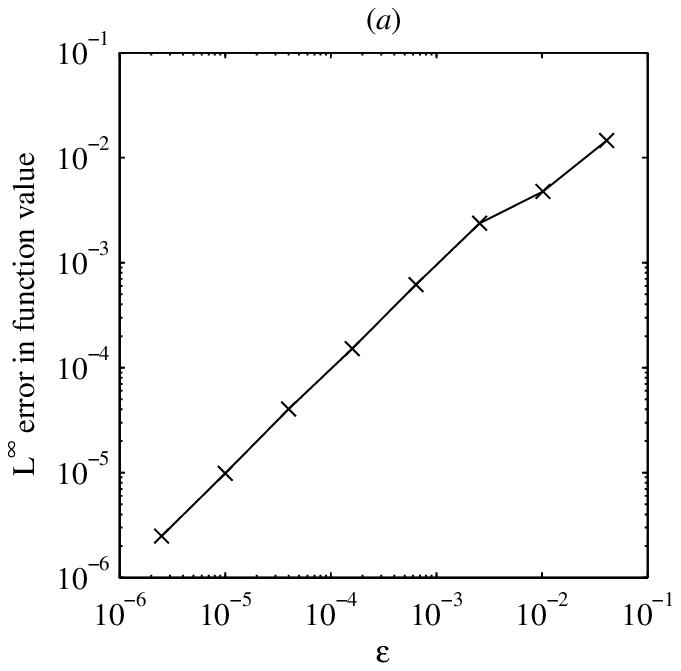}
\quad
\includegraphics[scale=0.8]{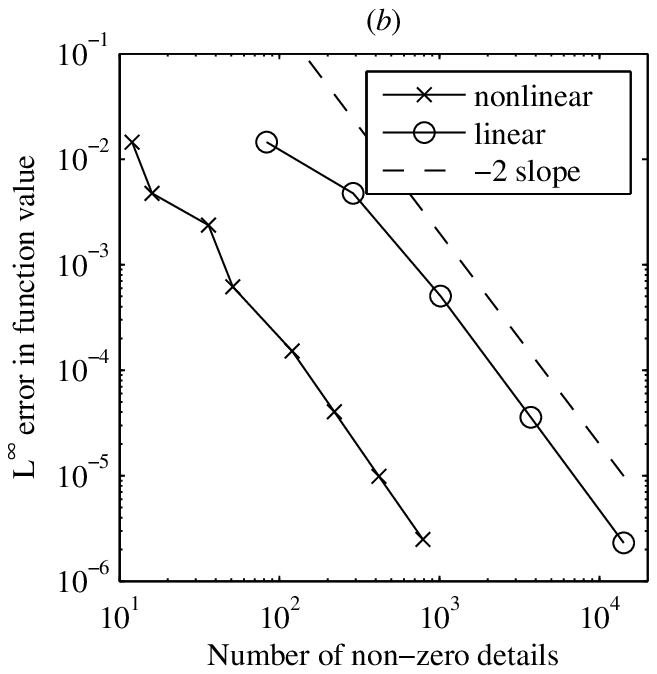}
\caption{(\textit{a}) Decay of $L^\infty$ error between the exact and reconstructed function values of a periodized two-dimensional Gaussian hump (\ref{eq:gaussian_mra}) with $\alpha=0.05$; (\textit{b}) decay of the $L^\infty$ error
versus the number of details retained by nonlinear (\ref{eq:cutoff_scind}) and linear filtering.}\label{fig:gauss_mra_errlinf}
\end{figure}

Figure~\ref{fig:chirp1d_a256_errlinf}(\textit{a}) confirms that the nonlinear filtering error scales like $\varepsilon$.
The error was estimated on a grid with $8192$ points.
The error versus the number of retained details $N_{details}$ is shown in figure~\ref{fig:chirp1d_a256_errlinf}(\textit{b}) for the linear and for the nonlinear filtering.
Note that, since this is a one-dimensional test, the number of details retained after the linear filtering
is in inverse proportion to the finest-level grid step size.
The figure shows that the error scales like $\mathcal{O}(N_{details}^{-4})$ in both cases.
However, for the same value of $N_{details}$, the error of nonlinear filtering is one order of magnitude smaller than its linear counterpart.

Let us now consider a two-dimensional example.
Let the function $u$ be a Gaussian, periodized (approximately) in order to conform the boundary conditions,
\begin{equation}
\begin{array}{l}
\displaystyle u(x_1,x_2) = \sum_{i_1=-p}^{p} \sum_{i_2=-p}^{p} g(x_1-i_1,x_2-i_2), \quad \textrm{where } \\
\displaystyle g(x_1,x_2) = \exp\left( - ((x_1-0.5)^2+(x_2-0.75)^2) / r_0^2 \right), \quad p=30.
\end{array}
\label{eq:gaussian_mra}
\end{equation}
In this example, we assign $r_0=0.05$.
Figure~\ref{fig:gauss_mra_errlinf}(\textit{a}) displays the same scaling of the nonlinear filtering error with $\varepsilon$ as in the previous example. Figure~\ref{fig:gauss_mra_errlinf}(\textit{b}) compares the linear and the nonlinear filtering. Now the number of details retained by the linear filtering is inversely proportional to the square of finest-level grid step size, therefore the error scales like $\mathcal{O}(N_{details}^{-2})$.
This figure also suggests that a Gaussian with $r_0=0.05$ is sufficiently well localised for the nonlinear filtering to be efficient. For the same precision, it results in 10 to 20 times less non-zero details than the linear filtering.

\subsection{Mixing of a periodized Gaussian hump}\label{sec:mixing_example}

A classical benchmark for level-set methods is the deformation of a contour with an unsteady velocity field.
Here we revisit the swirl test described in \cite{Seibold_etal_2012}.
In this section, we only discuss one example of an adaptive computation.
More detailed convergence and performance tests are presented in the following sections.

The initial condition $u_0(x_1,x_2)$ is a periodized Gaussian (\ref{eq:gaussian_mra}) with $r_0^2=0.1$.
The velocity field is given by
\begin{equation}
\bm{a}(x_1,x_2,t) = \cos\left(\frac{\pi t}{t_a}\right) \left(
\begin{array}{c}
\sin^2(\pi x_1) \sin(2 \pi x_2) \\
- \sin(2 \pi x_1) \sin^2(\pi x_2)
\end{array}
\right),
\label{eq:test_velocity}
\end{equation}
where $t_a = 10$. Note that this field is divergence-free, \ie~ $\nabla \cdot {\bm{a}}=0$.
The computational domain in space is a unit square and the time span is $T = 10$.
A computation has been carried out with threshold $\varepsilon = 5 \cdot 10^{-3}$.

\begin{figure}[htb]
\centering
\includegraphics[scale=0.75]{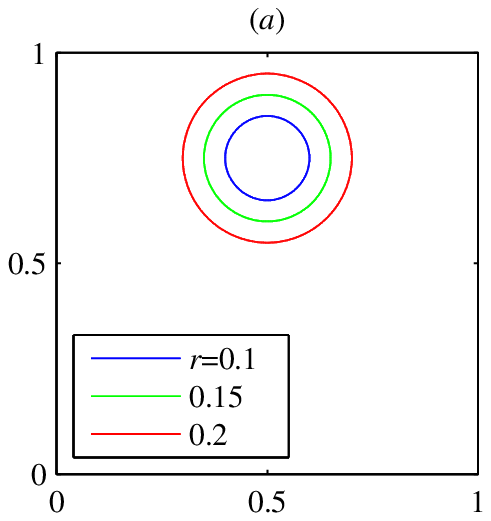}
~
\includegraphics[scale=0.75]{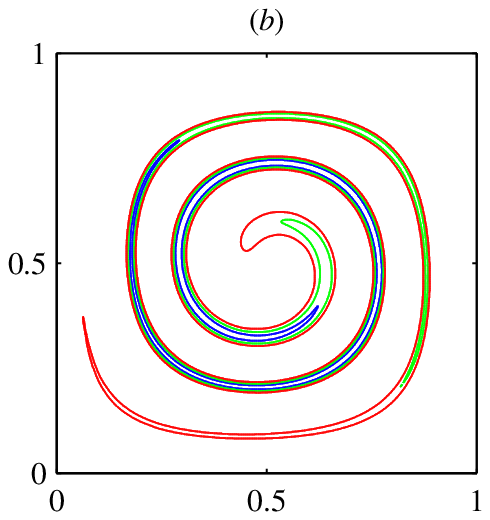}
~
\includegraphics[scale=0.75]{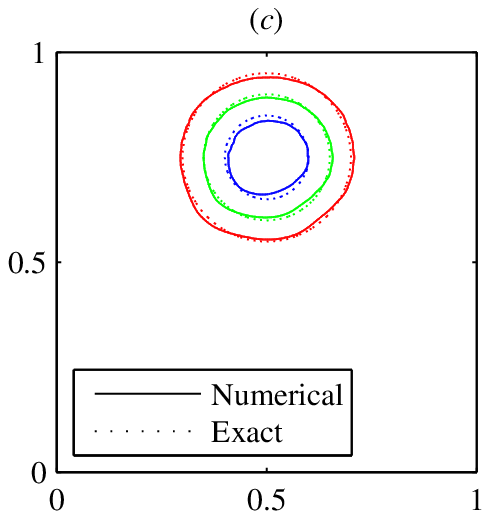}
\caption{(Colour online) Swirl test. Snapshots of isocontours $u=u_c(r)$ at time instants (\textit{a}) $t=0$, (\textit{b}) $t=5$ and (\textit{c}) $t=10$.
Continuous lines correspond to the numerical solution, dotted lines show the exact solution at $t=10$.
Red, green and blue are isolines $r=0.1$, $0.15$ and $0.2$, respectively.}\label{fig:snap_iso}
\end{figure}

Figure~\ref{fig:snap_iso} displays deformation of isocontours $u_c(r)=\exp(-10r^2)$ where $r=0.1$, 0.15 and 0.2, at time instants $t=0$, 5 and 10.
The initial almost circular contours wrap around the centre of the domain.
Maximum deformation occurs at $t=5$. Then the flow direction is reversed and the contours unwrap and restore their initial shape.
Thus the exact solution of the equation at $t=10$ coincides with the initial condition.
The numerical solution slightly differs from it.
The $L^\infty$ error in the function value is equal to 0.044.
The shape of the contours changes noticeably
due to numerical dispersion. The area enclosed by contours $r=0.2$ and 0.15 changes very little.
However, the method does not perfectly conserve the area (it should be conserved by the exact solution, since $\nabla \cdot \bm{a} = 0$);
this is seen in the decreased area of contour $r=0.1$, which is nearer to the maximum point.

\begin{figure}[htb]
\centering
\includegraphics[scale=0.75]{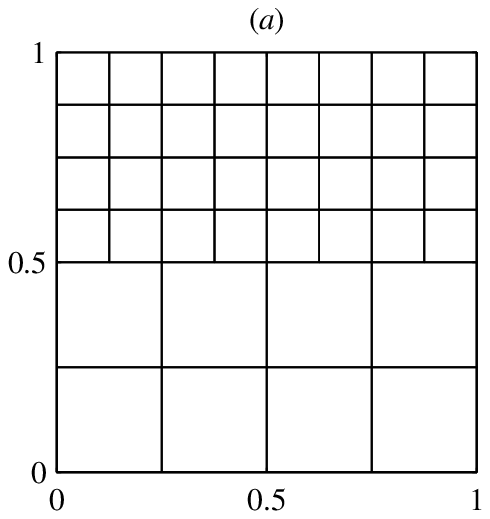}
~
\includegraphics[scale=0.75]{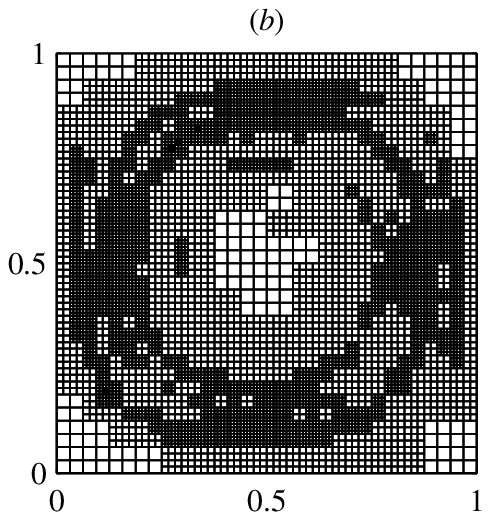}
~
\includegraphics[scale=0.75]{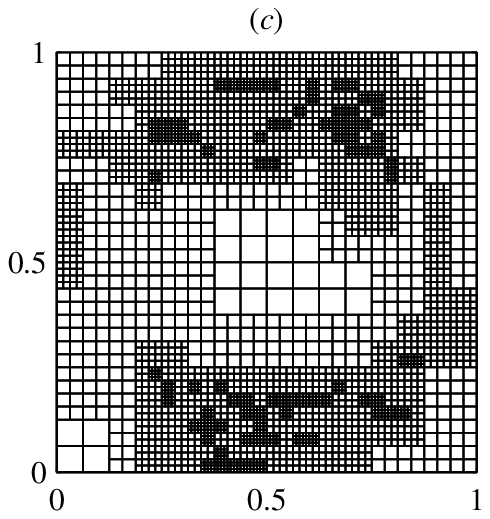}
\caption{Swirl test. Adaptive mesh at time instants (\textit{a}) $t=0$, (\textit{b}) $t=5$ and (\textit{c}) $t=10$.}\label{fig:snap_mesh}
\end{figure}

\begin{figure}[htb]
\centering
\includegraphics[scale=0.8]{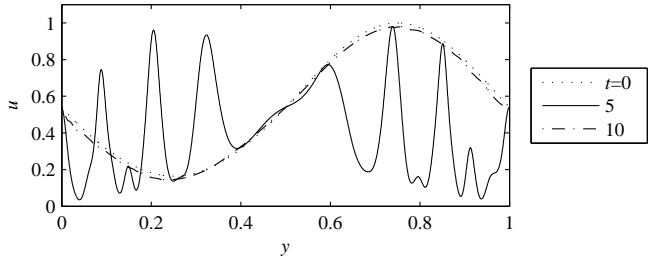}
\caption{Swirl test. The numerical solution $u$ sampled at the vertical line $x=0.5$ at time instants $t=0$ (initial condition), $t=5$ and $t=10$.}\label{fig:cuts_adpt}
\end{figure}

The adaptive mesh at the corresponding time instants is shown in figure~\ref{fig:snap_mesh}.
Multiresolution analysis of the initial condition produces a grid refined at the hump.
By the time $t=5$, the grid is refined everywhere in the domain. The most refinement occurs in an annulus around the centre of the domain,
where deformation is the strongest. Note that figure~\ref{fig:snap_iso} only displays three isocontours,
but the solution is defined in the whole domain and small scale features appear at multiple locations (see figure~\ref{fig:cuts_adpt}).
These locations are tracked by the mesh refinement algorithm.
At $t=10$ the grid coarsens.
However it is finer than the original one at $t=0$, and non-uniform.
Because of low diffusion and some dispersion of the scheme, the numerical solution accumulates spurious small scales.
Consequently, the grid has to be refined to capture these small features.
The finest cells in this computation correspond to level $l=8$, whereas the maximum allowed level is $M=11$, at which the solution plotted in figure~\ref{fig:snap_iso} is sampled.
Note that the details coefficients at levels $M+1$ and higher cannot be stored and they are assumed to be zero.
Therefore, if $M$ is fixed, the filtering becomes linear as $\varepsilon \to 0$.

\begin{figure}[htb]
\centering
\includegraphics[scale=0.8]{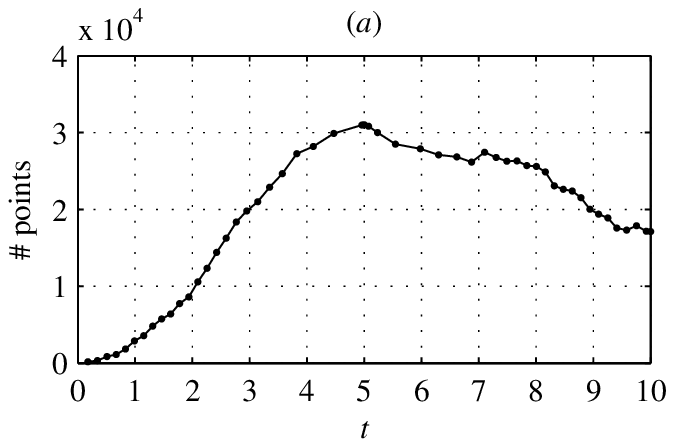}
\includegraphics[scale=0.8]{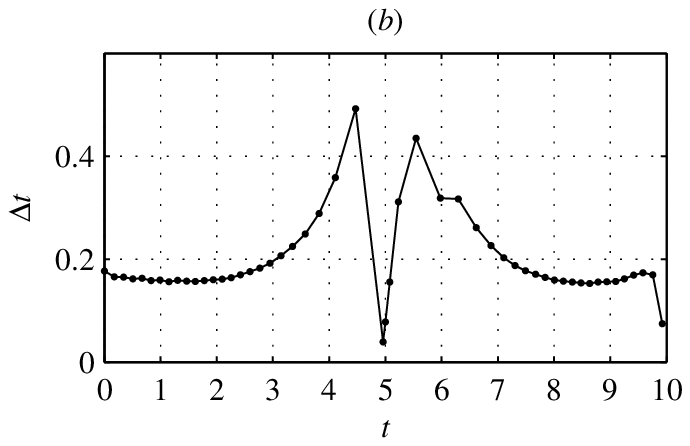}
\caption{Swirl test. (\textit{a}) Number of grid points and (\textit{b}) time step size $\Delta t$ versus time $t$. Markers indicate discrete time $t_n$.}\label{fig:numpoints}
\end{figure}

Figure~\ref{fig:numpoints}(\textit{a}) displays the time evolution of the number of grid points over time $t$.
The number of grid points $N_{points}$ is related to the number of quadtree nodes $N_{nodes}$:
\begin{equation}
N_{points} = 3 N_{nodes} + 1.
\end{equation}
It varies between 160 and 31024 in this simulation.
At the final step, the number of grid points equals 17104.
The total number of time steps is equal to 52.

The evolution of time step size $\Delta t$ is shown in figure~\ref{fig:numpoints}(\textit{b}).
During the first quarter-time, it is almost constant.
Then it increases as the velocity goes to zero.
However, at the time instant preceding $t=5$ (which is automatically detected by the algorithm), it is necessary to correct $\Delta t$
in order to produce the isolines shown in figure~\ref{fig:snap_iso}(\textit{b}).
Then $\Delta t$ grows geometrically, because of a limiter in (\ref{eq:new_time_step_size}),
until it restores to about the same size as before $t=5$.
Later, it decreases as the velocity increases,
and returns to about the same size as in the beginning of the computation.
Finally, one time step before the end of computation, $\Delta t$ is again corrected to produce output at exactly $t=10$.

\subsection{Convergence test on constant non-uniform grids}

\begin{figure}[htb]
\centering
\includegraphics[scale=0.8]{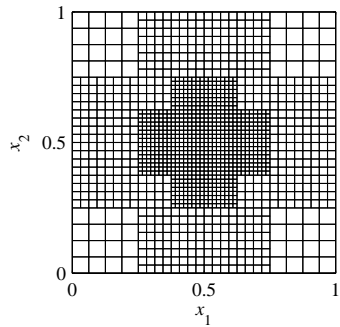}
\caption{Non-uniform grid used for convergence tests. By refining it uniformly, a series of grids was obtained.}\label{fig:fixed_mesh}
\end{figure}

Let us consider the non-uniform grid shown in figure~\ref{fig:fixed_mesh}, which has three levels only.
It is obtained from multiresolution analysis of a Gaussian function centered in the domain.
A finer grid can be constructed by splitting every cell into four finer cells. By repeating this process $K$ times, a series of $K$ grids is obtained.
Each of them has the same difference between the finest and the coarsest levels: it is equal to 2.

A series of computations on six different grids has been carried out,
with initial condition
\begin{equation}
u_0(x_1,x_2) = \cos(2 \pi x_1) \cos(4 \pi x_2)
\label{eq:inco_conv}
\end{equation}
and the velocity field given by (\ref{eq:test_velocity}) with $t_a=1$.
We recall that the exact solution of the problem coincides with the initial condition at time instants equal to multiples of $t_a$. Again, the adaptive time stepping method described in section~\ref{sec:time_stepping} has been employed.

\begin{figure}[htb]
\centering
\includegraphics[scale=0.8]{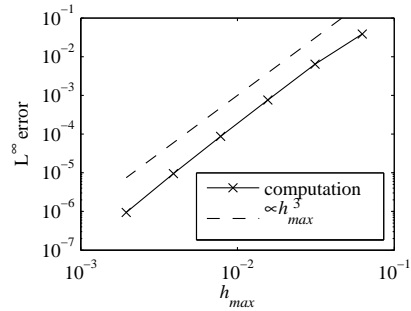}
\caption{Convergence test on constant non-uniform grids. Decay of the $L^\infty$ error versus $h_{max}$.}\label{fig:conv_fixed}
\end{figure}

The $L^\infty$ error between the numerical solution at $t=1$ and the exact solution has been calculated and plotted versus the maximum grid step size in each computation, $h_{max}$.
The result is displayed in figure~\ref{fig:conv_fixed}.
For intermediate values of $h_{max}$, the convergence is slightly faster than $\mathcal{O}(h_{max}^3)$ because the time discretization error decays faster and because of the adaptive time stepping.
Overall, these tests indicate the third-order rate of global convergence.
This is consistent with the $\mathcal{O}(h^4)$ local truncation error of the spatial discretization scheme
which yields a $\mathcal{O}(h^3)$ rate of decay of the global error on fixed uniform grids, as reported earlier \cite{Seibold_etal_2012}.

\subsection{Convergence tests on adaptive grids}\label{sec:conv_adaptive}

Convergence of the gradient-augmented level-set method on uniform non-adaptive grids was studied in \cite{Chidyagwal_etal_2012}.
In this section, we consider a series of adaptive computations with different threshold values $\varepsilon$, to analyze its influence.
The initial condition is again given by (\ref{eq:inco_conv}) and the velocity field is given by (\ref{eq:test_velocity}) with $t_a=1$.
In the simulations presented in this section, the maximum allowed
level is set to $M=15$. Note that, in practice, all grids generated during these computations were coarser:
the smallest grid step size $h_{min}$ in the most precise computation corresponded to $l=13$. I.e., no grid saturation occurred.

\begin{figure}[htb]
\centering
\includegraphics[scale=0.75]{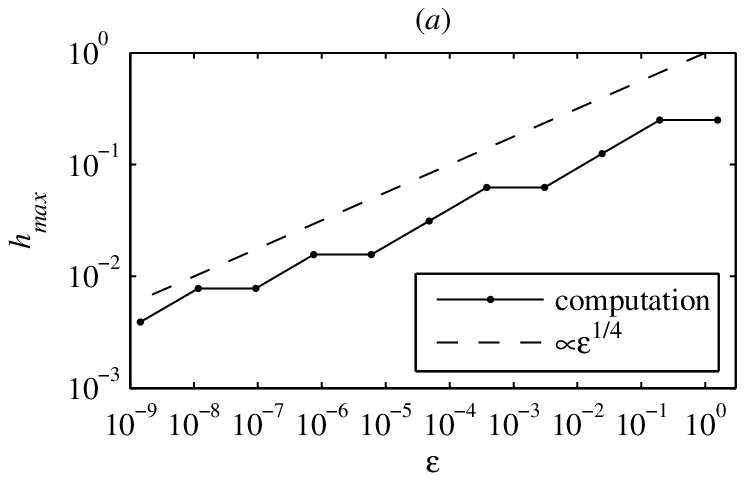}
\includegraphics[scale=0.75]{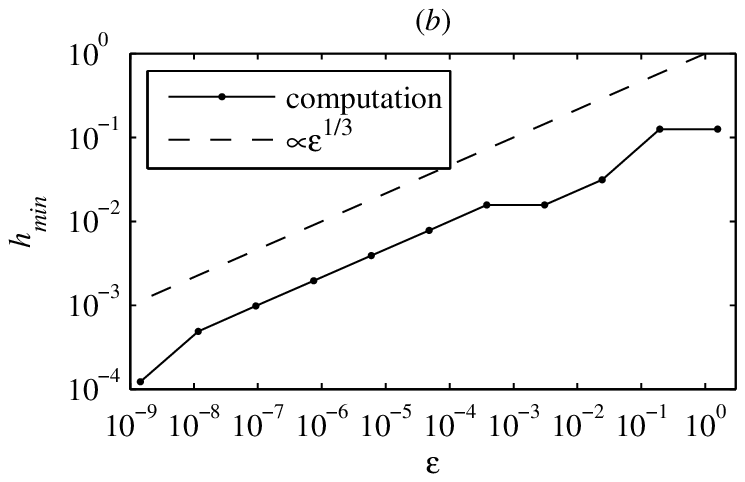}
\caption{Convergence tests. Decay of (\textit{a}) $h_{max}$ and (\textit{b}) $h_{min}$ versus $\varepsilon$.}\label{fig:adaptive_hminhmax}
\end{figure}

Smaller values of $\varepsilon$ result in finer grids. 
Let us define $h_{max}$ - the maximum cell size in the computation and $h_{min}$ - the minimum cell size.
The maximum and minimum are taken over all time steps.
Figure~\ref{fig:adaptive_hminhmax} confirms that both $h_{max}$ and $h_{min}$ become smaller with decreasing $\varepsilon$.
The rate is, however, faster for $h_{min}$ than for $h_{max}$.
In fact, $h_{max}$ in these tests typically corresponds to the largest cell of the mesh at $t=0$, which is obtained by the multiresolution analysis of the initial condition.
Therefore it asymptotically scales like $\varepsilon^{1/4}$.
The minimum cell size $h_{min}$ scales like $\varepsilon^{1/3}$, maybe because of accumulation of local small-scale errors.

\begin{figure}[htb]
\centering
\includegraphics[scale=0.75]{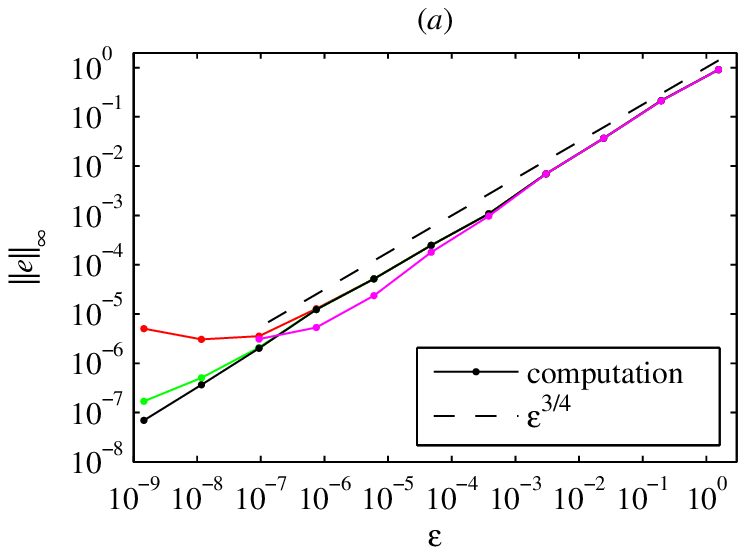}
\includegraphics[scale=0.75]{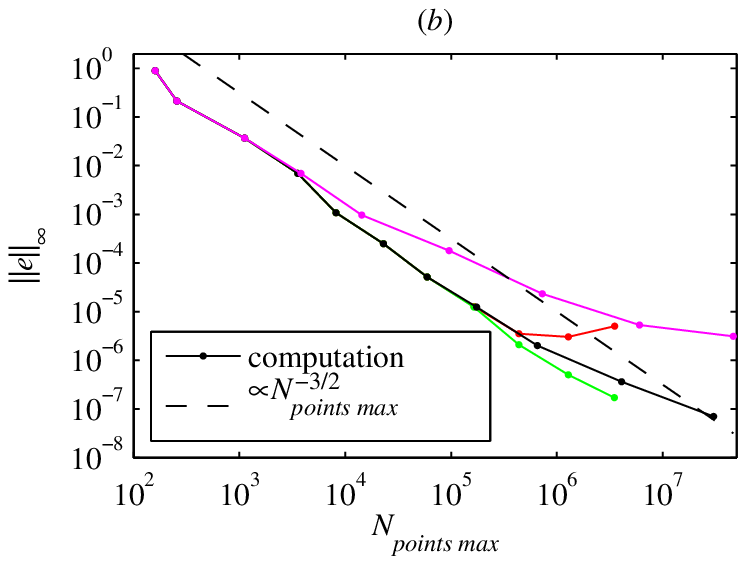}
\caption{Convergence tests. Decay of the $L^\infty$ error (\textit{a}) versus $\varepsilon$ and (\textit{b}) versus the maximum number of grid points $N_{points~max}$.
Different colours correspond to different values of $\eta$: $\eta=10^{-5}$ (red), $10^{-6}$ (green), $10^{-7}$ (black), $10^{-8}$ (magenta).}\label{fig:adaptive_conv}
\end{figure}

The accuracy has been measured in the $L^\infty$ error norm between the solution at $t=1$ and the initial condition,
calculated over the grid-point values at the last step of each simulation.
Figure~\ref{fig:adaptive_conv}(\textit{a}) shows the decay of that error, $||e||_\infty$, versus $\varepsilon$.
Unlike in the multiresolution transform tests in section~\ref{sec:multi_validation}, the decay is slower than linear.
This can be briefly explained as follows.

Let us consider the local error introduced for a single time step of the method, defined as $e_{loc}=u(\bm{x},t_{n+1})-u_{n+1}$,
where $u(\bm{x},t_{n+1})$ is the exact solution at point $\bm{x}$ and time $t_{n+1}=t_n+\Delta t$, and $u_{n+1}$ is the numerical solution
obtained after a single time step starting from time $t_n$ and using $u_n=u(\bm{x},t_{n})$ as the initial condition.
By construction, the time integration of the characteristic equation introduces an error of order $\varepsilon$ at a single time step, as explained in section~\ref{sec:time_stepping}.
The interpolation in space also introduces an error of order $\varepsilon$. This is ensured by the nonlinear filtering of the multiresolution decomposition discussed in sections~\ref{sec:multiresolution} and \ref{sec:multi_validation}.
Hence, the local error scales like
\begin{equation}
e_{loc} \sim \varepsilon.
\end{equation}
The global error can be approximately estimated as $||e||_\infty \sim N_{time~steps} ||e_{loc}||_\infty$,
where the number of time steps $N_{time~steps}$ depends on the adaptive time step size $\Delta t$.
The time step size control assumes that the local truncation error of the scheme is $\mathcal{O}(\Delta t^4)$, therefore $\Delta t = \mathcal{O}(\varepsilon^{1/4})$.
Then $N_{time~steps} = \mathcal{O}(\varepsilon^{-1/4})$ and we obtain
\begin{equation}
||e||_\infty = \mathcal{O}(\varepsilon^{3/4}).
\end{equation}
This estimate agrees well with the numerical results, as shown in figure~\ref{fig:adaptive_conv}(\textit{a}).
Note that the error constant is of order unity.

Another important scaling is displayed in figure~\ref{fig:adaptive_conv}(\textit{b}).
It depicts the decay of $||e||_\infty$ versus $N_{points~max}$, the maximum number of grid points achieved in an adaptive simulation.
For comparison, in a fixed uniform grid computation, the number of grid points does not change over time and it is proportional to $1/h^2$ (for a two-dimensional domain).
Since the numerical method is globally third-order accurate, we obtain an estimate
\begin{equation}
||e||_\infty = \mathcal{O}(N_{points~max}^{-3/2})
\end{equation}
that appears to hold not only for fixed-grid, but also for adaptive computations, if local refinement is moderate (\ie, when $h_{max}/h_{min}<16$, in this example).
Thus, the order of the discretization scheme is maintained by the adaptive method.


Further, figure~\ref{fig:adaptive_conv} presents a test of sensitivity to the choice of finite-difference parameter $\eta$.
In addition to the `default' value $\eta=10^{-7}$, similar computations have been carried out with $\eta=10^{-5}$, $10^{-6}$ and $10^{-8}$.
The $L^\infty$ error and the maximum number of grid points are both sensitive to $\eta$, i.e., when $\varepsilon$ is small,
the error due to the finite-difference approximation of $u_{x_{1}}$, $u_{x_{2}}$, $u_{x_{1}x_{2}}$
becomes dominant. In the case of $\eta=10^{-5}$, this is essentially the truncation error.
As $\varepsilon$ decreases, $||e||_\infty$ first decreases, then saturates at $||e||_\infty \approx 3\cdot10^{-6}$.
In the case of $\eta=10^{-8}$, however, the round-off errors become dominant. The $L^\infty$ error again saturates at the level of $||e||_\infty \approx 3\cdot10^{-6}$,
and, in addition, the maximum number of grid points for a given $\varepsilon$ increases.
As discussed in \cite{Chidyagwal_etal_2012}, the round-off errors incurred by the finite-difference scheme are of magnitude $\mathcal{O}(\delta^{1/2})$,
where $\delta$ is the accuracy of the floating point operations.
The values $\eta=10^{-6}$...$10^{-7}$ allow reaching the error as small as $10^{-7}$.
Therefore, all computations further in this paper have been carried out with $\eta=10^{-7}$.

Since the finite-difference approximation is second-order accurate, and
the gradient-augmented scheme is third-order accurate, it is possible to estimate the minimum
allowed cell size as $h_{min~inf}=\varkappa \eta^{2/3}$, where $\varkappa=\mathcal{O}(1)$,
supposing that the error constants of both methods are of the same order of magnitude.
For $\eta=10^{-7}$ we obtain $h_{min~inf}=2\cdot10^{-5}$, which can
be reached in a computation with $M=15$ levels.

\begin{figure}[htb]
\centering
\includegraphics[scale=0.75]{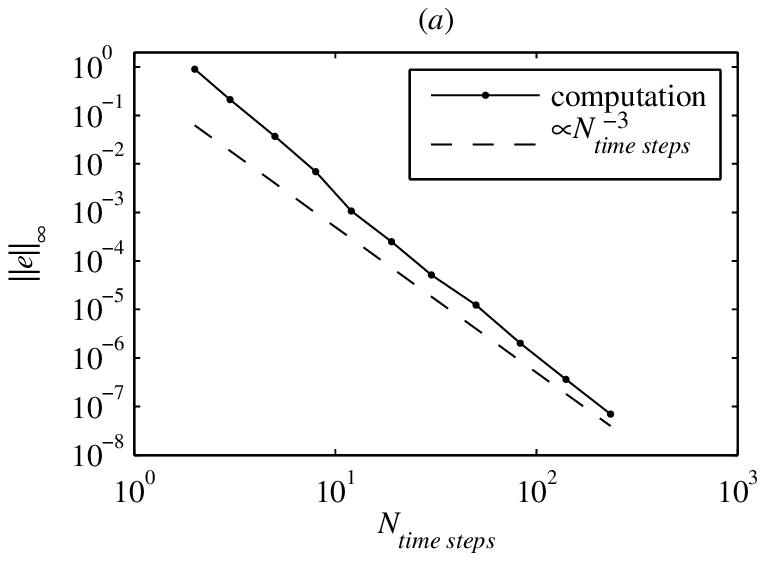}
\includegraphics[scale=0.75]{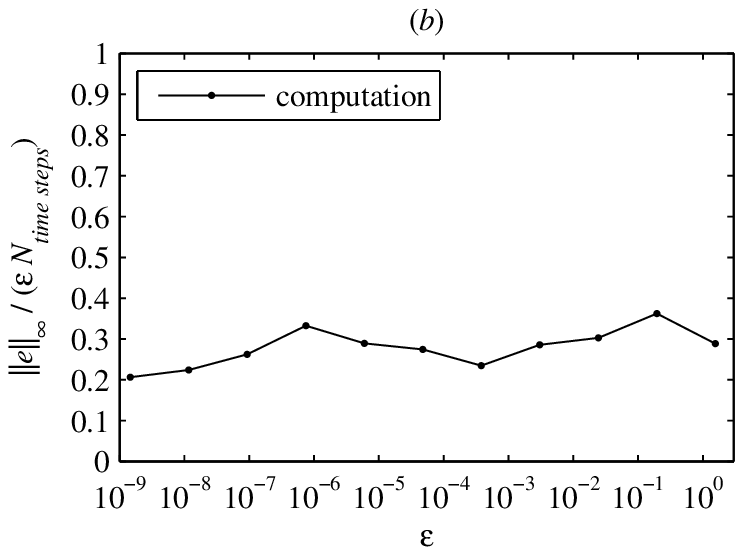}
\caption{Convergence tests. (\textit{a}) Decay of the $L^\infty$ error versus number of time steps $N_{time~steps}$.
(\textit{b}) Ratio of the $L^\infty$ error over its estimate $\varepsilon N_{time~steps}$, versus $\varepsilon$. Time $t=1$.}\label{fig:adaptive_conv_timestep}
\end{figure}

The decay rate of the $L^\infty$ error versus number of time steps is consistent with the third-order error estimate formula used for time step size control.
This is shown in figure~\ref{fig:adaptive_conv_timestep}(\textit{a}). Assuming that the local error at every time step is of order $\varepsilon$,
the $L^\infty$ error can be estimated as $\varepsilon N_{time~steps}$. The ratio between the actual error and this estimate is plotted in figure~\ref{fig:adaptive_conv_timestep}(\textit{b}).
It oscillates between 0.2 and 0.4 without any significant trend. This suggests that the adaptive method controls the local error as desired.

Using the scalings obtained in these convergence tests,
it is possible to devise a method for computing with a given tolerance $tol \approx ||e||_\infty$ at $t=T$.
First, a preliminary computation is carried out with $\varepsilon=\varepsilon_1$, where $\varepsilon_1$ is relatively large such that the computation is reasonably fast.
The number of time steps in this computation is $N_{time~steps~1}$. Then the final computation is carried out with $\varepsilon=\varepsilon_2$, where
\begin{equation}
\varepsilon_2 = \left( \frac{tol~~\varepsilon_1^4}{C~~N_{time~steps~1}} \right)^{1/5}
\end{equation}
with $C \approx 0.3$ (see figure~\ref{fig:adaptive_conv_timestep}\textit{b}).

\subsection{Performance tests}\label{sec:performance}

In this section, we compare computational cost of adaptive and uniform fixed-grid simulations.
The initial condition is again a periodized Gaussian (\ref{eq:gaussian_mra}) of radius $r_0$.
A series of adaptive simulations has been carried out with different values of $r_0$.
The velocity field is given by (\ref{eq:test_velocity}) with $t_a=3$, and the simulations are stopped at $t=3$.
Note that we chose $r_0=\sqrt{0.1}\approx0.316$ for the initial condition of the swirl test described in section~\ref{sec:mixing_example}.

\begin{figure}[htb]
\centering
\includegraphics[scale=0.75]{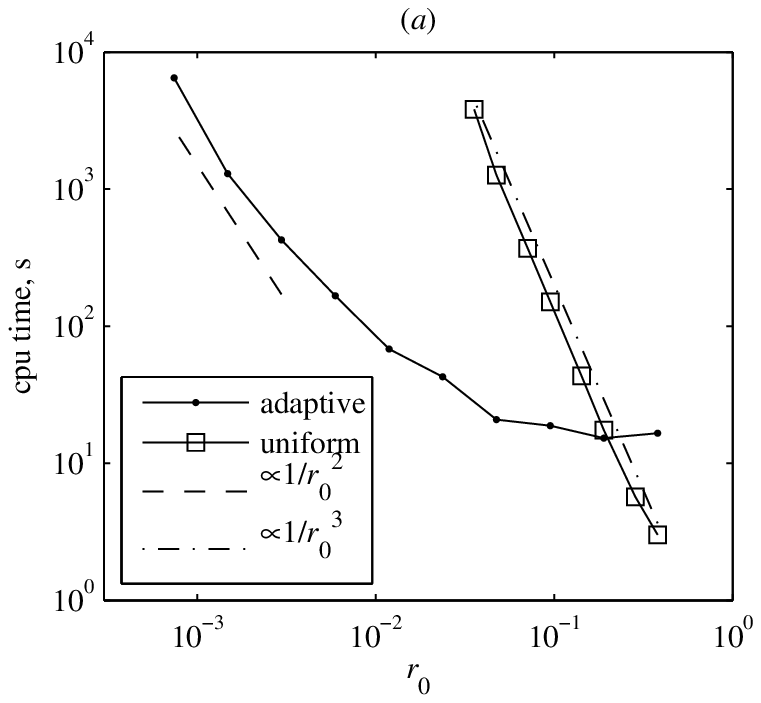}
\includegraphics[scale=0.75]{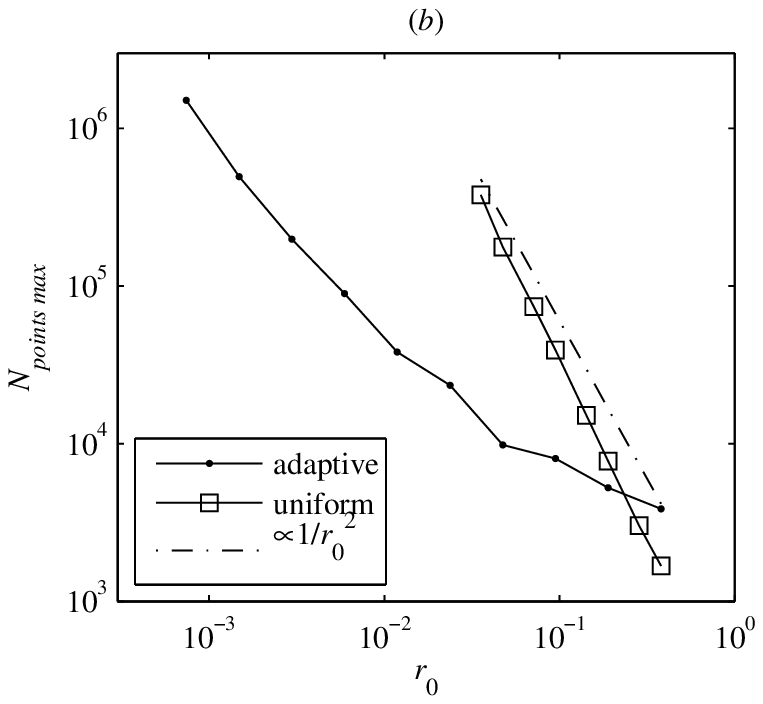}
\caption{Performance tests. (\textit{a}) CPU time, in seconds, and (\textit{b}) number of grid points $N_{points~max}$, versus $r_0$.}\label{fig:adaptive_timing}
\end{figure}

In each adaptive simulation, the threshold $\varepsilon$ was set to an appropriate value such that
the $L^\infty$ error at $t=3$ was equal to $0.01 \pm 0.0004$.
The minimum possible level was set to $m=4$.
CPU time and maximum number of grid points (representative of memory usage) were measured.
They are displayed in figures~\ref{fig:adaptive_timing}(\textit{a}) and (\textit{b}), respectively.

For comparison, similar computations have been carried out using a different code,
which implements the numerical method described in section~\ref{sec:levelset_method} on uniform Cartesian grids (see the original reference \cite{Seibold_etal_2012}).
These results are also shown in figure~\ref{fig:adaptive_timing}.

If $r_0$ is large, the adaptive code is slower than the non-adaptive one.
Note that, if the adaptive code is constrained to uniform space and time discretization, it is about 7 times slower than the specialized uniform-grid code.
This factor partly consists of the cost of error estimation, which requires one extra level of grid refinement. If a uniform grid consists of $N_{nodes}$ cells,
the adaptive algorithm operates $3N_{nodes}+1$ points, while the non-adaptive algorithm operates only $N_{nodes}$ grid points.
In addition, some overhead is due to the tree data structure, because it requires $\mathcal{O}(\log N_{nodes})$ operations every time a grid point value is accessed.
In the present tests, the adaptive grid is, in general, not uniform. Therefore, the adaptive code is, at worst, only 5 times slower than the non-adaptive.

At $r_0 \approx 0.2$, both codes have equal efficiency.
The CPU time of the non-adaptive computations scales like $1/r_0^3$ whereas
adaptivity reduces it drastically.
The parameter $r_0$ controls the size of the Gaussian hump.
As it decreases, the grid has to be refined to ensure the desired accuracy.
However, refinement is only necessary in a small circle of radius of order $r_0$.
Therefore, the number of grid points in the adaptive computations depends more weakly on $r_0$ than in the fixed-grid computations.
The number of time steps is, at worst, in inverse proportion to $h_{min} \propto r_0$.
The CPU time in the adaptive computations shows two regimes. It is nearly constant if $r_0$ is large, and it scales like $1/r_0^2$ if $r_0$ is small.

In contrast, if the grid is uniform, the number of grid points required for a given accuracy behaves like $1/r_0^2$, since the grid step $h$ should be proportional to $r_0$.
If the time step is equal to the grid step size $h$, the computational complexity increases like $1/r_0^3$ as $r_0 \to 0$.
Therefore, eventually, the adaptive code outperforms the non-adaptive one when $r_0$ is sufficiently small.
We anticipate the same behaviour of this method applied to other problems that focus on time evolution of well localized features of point-singularity type.

\section{Conclusions and perspectives}\label{sec:conclusions}

A semi-Lagrangian adaptive numerical method has been developed for the two-dimensional advection equation.
It is based on the gradient-augmented level set method \cite{Nave_etal_2010}.
This numerical scheme uses Hermite interpolation. It is compact, third order accurate, and unconditionally stable.
Multiresolution decomposition is employed to obtain an error estimate required for adaptivity in space, while an embedded Runge--Kutta scheme is applied for the time discretization error estimate and for adapting automatically the time step.
For consistency with the gradient-augmented method, the multiresolution scheme is based on Hermite interpolation \cite{Warming_Beam_2000}.
Its implementation uses quadtree data structures with dynamic memory allocation.

The $L^\infty$ error norm of the numerical solution is controlled by $\varepsilon$ - threshold of the multiresolution scheme.
Numerical experiments suggest that the error scales like $\varepsilon^{3/4}$, which we justified by heuristic arguments.

A series of numerical experiments has been carried out with advection of a Gaussian hump.
The size of its support $r_0$ has been varied.
These experiments show that, the more localized the hump is, the more beneficial the adaptive method becomes when compared to the uniform discretization approach, in terms of CPU time and memory compression.

The originality of the current work is the coupling of the gradient-augmented level set method
with an adaptive multiresolution method and adaptive time stepping. This allows for speed up of CPU time
and memory compression, i.e., the new adaptive solver is more efficient than the one on regular grids and in addition
the errors in space and time are controlled. The order of the underlying discretization scheme on a regular grid is maintained.


It is straightforward to generalize the method to three-dimensional problems.
A possible generalization to the incompressible Euler equation is of interest.
There are possible applications in plasma physics, e.g., a further improvement of the particle-in-wavelet method for the {V}lasov--{P}oisson equations \cite{Nguyen_van_yen_etal_2011}.
Finally, we anticipate using similar techniques for solving elasticity equations.

\begin{acknowledgements}
DK acknowledges financial support from the CRM--ISM Fellowship and thanks Alexey Eremin for useful discussions about Runge--Kutta schemes.
JCN acknowledges support from the NSERC Discovery and Discovery Accelerator Programs.
KS thankfully acknowledges financial support from the ANR project SiCoMHD (ANR-Blanc 2011-045).
\end{acknowledgements}

\appendix
\section{Two-dimensional Hermite interpolation}\label{sec:appendix}

\begin{figure}[htb]
\centering
\includegraphics[scale=0.8]{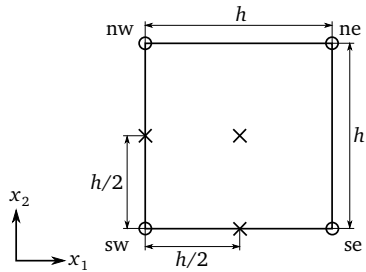}
\caption{Interpolation cell.
Markers $\circ$ show the corner points, where the values of the function $u$ and its derivatives $u_{x_1}$, $u_{x_2}$ and $u_{x_1 x_2}$ are known.
Markers $\times$ show the 3 points that appear in mid-point interpolation formulae (\ref{eq:mid_u})-(\ref{eq:mid_ux1x2}).
}\label{fig:drawing_appendix}
\end{figure}

Suppose that values of a function $u(x_1,x_2)$ and its derivatives
$u_{x_1}$, $u_{x_2}$ and $u_{x_1 x_2}$
are given at four vertices of a square of side $h$, as shown in figure~\ref{fig:drawing_appendix}.
We will use superscripts $sw$, $se$, $nw$ and $ne$ to refer to these points and the corresponding values.

Let us rescale the coordinates $x_1$, $x_2$:
\begin{equation}
 \tilde{x}_1 = \frac{x_1-x^{\textrm{sw}}_1}{h}, \quad\quad \tilde{x}_2 = \frac{x_2-x^{\textrm{sw}}_2}{h},
\end{equation}
and define basis functions
\begin{equation}
 f(\tilde{x}) = 2\tilde{x}^3-3\tilde{x}^2+1, \quad\quad g(\tilde{x}) = \tilde{x}^3-2\tilde{x}^2+\tilde{x}
\label{eq:hermite_basis}
\end{equation}
The value of $u$ at $(x_1,x_2) \in [x^{\textrm{sw}}_1,x^{\textrm{sw}}_1+h]\times[x^{\textrm{sw}}_2,x^{\textrm{sw}}_2+h]$
can be estimated using the following $\mathcal{O}(h^4)$ accurate formula:
\begin{equation}
\begin{array}{rcl}
 \tilde{u}(x_1,x_2) & = & ( u^{\textrm{sw}} f(\tilde{x}_1) f(\tilde{x}_2) + u^{\textrm{se}} f(1-\tilde{x}_1) f(\tilde{x}_2) \\
                    &   &     \quad\quad\quad\quad + u^{\textrm{nw}} f(\tilde{x}_1) f(1-\tilde{x}_2) + u^{\textrm{ne}} f(1-\tilde{x}_1) f(1-\tilde{x}_2) ) \\
                    &   & +h~  ( u_{x_1}^{\textrm{sw}} g(\tilde{x}_1) f(\tilde{x}_2) - u_{x_1}^{se} g(1-\tilde{x}_1) f(\tilde{x}_2) \\
                    &   &     \quad\quad\quad\quad + u_{x_1}^{\textrm{nw}} g(\tilde{x}_1) f(1-\tilde{x}_2) - u_{x_1}^{\textrm{ne}} g(1-\tilde{x}_1) f(1-\tilde{x}_2) ) \\
                    &   & +h~  ( u_{x_2}^{\textrm{sw}} f(\tilde{x}_1) g(\tilde{x}_2) + u_{x_2}^{\textrm{se}} f(1-\tilde{x}_1) g(\tilde{x}_2) \\
                    &   &     \quad\quad\quad\quad - u_{x_2}^{\textrm{nw}} f(\tilde{x}_1) g(1-\tilde{x}_2) - u_{x_2}^{\textrm{ne}} f(1-\tilde{x}_1) g(1-\tilde{x}_2) ) \\
                    &   & + h^2~  ( u_{x_1 x_2}^{\textrm{sw}} g(\tilde{x}_1) g(\tilde{x}_2) - u_{x_1 x_2}^{\textrm{se}} g(1-\tilde{x}_1) g(\tilde{x}_2) \\
                    &   &     \quad\quad\quad\quad - u_{x_1 x_2}^{\textrm{nw}} g(\tilde{x}_1) g(1-\tilde{x}_2) + u_{x_1 x_2}^{\textrm{ne}} g(1-\tilde{x}_1) g(1-\tilde{x}_2) ).
\end{array}
\label{eq:hermite_2d}
\end{equation}
It is straightforward to obtain interpolation formulae for the first and second partial derivatives of $u$ by derivating (\ref{eq:hermite_2d}).

The values $\tilde{u}(x^{\textrm{sw}}_1+\frac{h}{2},x^{\textrm{sw}}_2)$, $\tilde{u}(x^{\textrm{sw}}_1,x^{\textrm{sw}}_2+\frac{h}{2})$ and $\tilde{u}(x^{\textrm{sw}}_1+\frac{h}{2},x^{\textrm{sw}}_2+\frac{h}{2})$,
as well as unscaled derivatives
required for the error estimate,
are also obtained from (\ref{eq:hermite_2d}).
For multiresolution decomposition (\ref{eq:decomposition})
and reconstruction (\ref{eq:reconstruction}),
we define scaled quantities:
\begin{equation}
\begin{array}{llll}
u_0^{00} = u^{sw}, & u_0^{20} = u^{se}, & u_0^{02} = u^{nw}, & u_0^{22} = u^{ne}, \\
u_1^{00} = \frac{h}{2} u_{x_1}^{sw}, & u_1^{20} = \frac{h}{2} u_{x_1}^{se}, & u_1^{02} = \frac{h}{2} u_{x_1}^{nw}, & u_1^{22} = \frac{h}{2} u_{x_1}^{ne}, \\
u_2^{00} = \frac{h}{2} u_{x_2}^{sw}, & u_2^{20} = \frac{h}{2} u_{x_2}^{se}, & u_2^{02} = \frac{h}{2} u_{x_2}^{nw}, & u_2^{22} = \frac{h}{2} u_{x_2}^{ne}, \\
u_3^{00} = \frac{h^2}{4} u_{x_1 x_2}^{sw}, & u_3^{20} = \frac{h^2}{4} u_{x_1 x_2}^{se}, & u_3^{02} = \frac{h^2}{4} u_{x_1 x_2}^{nw}, & u_3^{22} = \frac{h^2}{4} u_{x_1 x_2}^{ne},
\end{array}
\label{eq:scaled_quantities_corner}
\end{equation}
as well as
\begin{equation}
\begin{array}{ll}
\tilde{u}_0^{10} = \tilde{u}(x^{\textrm{sw}}_1+\frac{h}{2},x^{\textrm{sw}}_2), & \tilde{u}_0^{01} = \tilde{u}(x^{\textrm{sw}}_1,x^{\textrm{sw}}_2+\frac{h}{2}), \\ & \tilde{u}_0^{11} = \tilde{u}(x^{\textrm{sw}}_1+\frac{h}{2},x^{\textrm{sw}}_2+\frac{h}{2}), \\
\tilde{u}_1^{10} = \frac{h}{2} \tilde{u}_{x_1}(x^{\textrm{sw}}_1+\frac{h}{2},x^{\textrm{sw}}_2), & \tilde{u}_1^{01} = \frac{h}{2} \tilde{u}_{x_1}(x^{\textrm{sw}}_1,x^{\textrm{sw}}_2+\frac{h}{2}), \\ & \tilde{u}_1^{11} = \frac{h}{2} \tilde{u}_{x_1}(x^{\textrm{sw}}_1+\frac{h}{2},x^{\textrm{sw}}_2+\frac{h}{2}), \\
\tilde{u}_2^{10} = \frac{h}{2} \tilde{u}_{x_2}(x^{\textrm{sw}}_1+\frac{h}{2},x^{\textrm{sw}}_2), & \tilde{u}_2^{01} = \frac{h}{2} \tilde{u}_{x_2}(x^{\textrm{sw}}_1,x^{\textrm{sw}}_2+\frac{h}{2}), \\ & \tilde{u}_2^{11} = \frac{h}{2} \tilde{u}_{x_2}(x^{\textrm{sw}}_1+\frac{h}{2},x^{\textrm{sw}}_2+\frac{h}{2}), \\
\tilde{u}_3^{10} = \frac{h^2}{4} \tilde{u}_{x_1 x_2}(x^{\textrm{sw}}_1+\frac{h}{2},x^{\textrm{sw}}_2), & \tilde{u}_3^{01} = \frac{h^2}{4} \tilde{u}_{x_1 x_2}(x^{\textrm{sw}}_1,x^{\textrm{sw}}_2+\frac{h}{2}), \\ & \tilde{u}_3^{11} = \frac{h^2}{4} \tilde{u}_{x_1 x_2}(x^{\textrm{sw}}_1+\frac{h}{2},x^{\textrm{sw}}_2+\frac{h}{2}).
\end{array}
\label{eq:scaled_quantities_mid}
\end{equation}
Thus we obtain the following formulae:
\begin{equation}
\begin{array}{rcl}
   \tilde{u}_0^{10} & = & \frac{1}{2} (u^{00}+u^{10}) + \frac{1}{4} (u_{x1}^{00}-u_{x1}^{10}), \\
   \tilde{u}_0^{01} & = & \frac{1}{2} (u^{00}+u^{01}) + \frac{1}{4} (u_{x2}^{00}-u_{x2}^{01}), \\
   \tilde{u}_0^{11} & = & \frac{1}{4} (u^{00}+u^{10}+u^{01}+u^{11}) \\
   & & + \frac{1}{8} \left( (u_{x1}^{00}-u_{x1}^{10}+u_{x1}^{01}-u_{x1}^{11}) + (u_{x2}^{00}+u_{x2}^{10}-u_{x2}^{01}-u_{x2}^{11}) \right) \\
   & & + \frac{1}{16} (u_{x_1 x_2}^{00}-u_{x_1 x_2}^{10}-u_{x_1 x_2}^{01}+u_{x_1 x_2}^{11}),
\end{array}
\label{eq:mid_u}
\end{equation}
\begin{equation}
\begin{array}{rcl}
   \tilde{u}_1^{10} & = & - \frac{3}{4} (u^{00}-u^{10}) - \frac{1}{4} (u_{x_1}^{00}+u_{x_1}^{10}), \\
   \tilde{u}_1^{01} & = & \frac{1}{2} (u_{x_1}^{00}+u_{x_1}^{01}) + \frac{1}{4} (u_{x_1 x_2}^{00}-u_{x_1 x_2}^{01}), \\
   \tilde{u}_1^{11} & = & \frac{3}{8} (-u^{00}+u^{10}-u^{01}+u^{11})  \\
   & & - \frac{1}{8} (u_{x_1}^{00}+u_{x_1}^{10}+u_{x_1}^{01}+u_{x_1}^{11}) - \frac{3}{16} (u_{x_2}^{00}-u_{x_2}^{10}-u_{x_2}^{01}+u_{x_2}^{11}) \\
   & & - \frac{1}{16} (u_{x_1 x_2}^{00}+u_{x_1 x_2}^{10}-u_{x_1 x_2}^{01}-u_{x_1 x_2}^{11})
\end{array}
\label{eq:mid_ux1}
\end{equation}
\begin{equation}
\begin{array}{rcl}
   \tilde{u}_2^{10} & = & \frac{1}{2} (u_{x_2}^{00}+u_{x_2}^{10}) + \frac{1}{4} (u_{x_1 x_2}^{00}-u_{x_1 x_2}^{10}), \\
   \tilde{u}_2^{01} & = & - \frac{3}{4} (u^{00}-u^{01}) - \frac{1}{4} (u_{x_2}^{00}+u_{x_2}^{01}), \\
   \tilde{u}_2^{11} & = & \frac{3}{8} (-u^{00}-u^{10}+u^{01}+u^{11}) \\
   & & - \frac{3}{16} (u_{x_1}^{00}-u_{x_1}^{10}-u_{x_1}^{01}+u_{x_1}^{11}) - \frac{1}{8} (u_{x_2}^{00}+u_{x_2}^{10}+u_{x_2}^{01}+u_{x_2}^{11}) \\
   & & - \frac{1}{16} (u_{x_1 x_2}^{00}-u_{x_1 x_2}^{10}+u_{x_1 x_2}^{01}-u_{x_1 x_2}^{11})
\end{array}
\label{eq:mid_ux2}
\end{equation}
\begin{equation}
\begin{array}{rcl}
   \tilde{u}_3^{10} & = & - \frac{3}{4} (u_{x_2}^{00}-u_{x_2}^{10}) - \frac{1}{4} (u_{x_1 x_2}^{00}+u_{x_1 x_2}^{10}), \\
   \tilde{u}_3^{01} & = & - \frac{3}{4} (u_{x_1}^{00}-u_{x_1}^{01}) - \frac{1}{4} (u_{x_1 x_2}^{00}+u_{x_1 x_2}^{01}), \\
   \tilde{u}_3^{11} & = & \frac{9}{16} (u^{00}-u^{10}-u^{01}+u^{11}) \\
   & & + \frac{3}{16} \left( (u_{x_1}^{00}+u_{x_1}^{10}-u_{x_1}^{01}-u_{x_1}^{11}) + (u_{x_2}^{00}-u_{x_2}^{10}+u_{x_2}^{01}-u_{x_2}^{11}) \right) \\
   & & + \frac{1}{16} (u_{x_1 x_2}^{00}+u_{x_1 x_2}^{10}+u_{x_1 x_2}^{01}+u_{x_1 x_2}^{11})
\end{array}
\label{eq:mid_ux1x2}
\end{equation}
In the notations of (\ref{eq:decomposition})-(\ref{eq:reconstruction}), we obtain
\begin{equation}
(\tilde{u}_\iota)_{2j_1+1,2j_2}^l = \tilde{u}_\iota^{10}, \quad
(\tilde{u}_\iota)_{2j_1,2j_2+1}^l = \tilde{u}_\iota^{01}, \quad
(\tilde{u}_\iota)_{2j_1+1,2j_2+1}^l = \tilde{u}_\iota^{11},
\end{equation}
where $\iota = 0,...,3$ and indices $j_1$, $j_2$ and $l$ are determined by $\bm{x}^{sw}$ and $h$, as described in sections~\ref{sec:multiresolution} and \ref{sec:tree}.
Note that the computational cost of this procedure is much less than interpolation at an arbitrary point because the coefficients that include the basis functions (\ref{eq:hermite_basis}) are pre-computed analytically.

\end{document}